%% file: main.tex
\documentclass[acmsmall,nonacm]{acmart}
\input{pluverse-latex-style-guide/macro/pluverse-preamble.tex}

\lstdefinelanguage{llvm}{
  morekeywords={define, declare, ret, call, void, ptr, global},
  morekeywords=[2]{add, sub, mul, shl, lshr, ashr, and, or, xor, icmp, zext, sext, load, store, getelementptr, select, br, phi},
  morekeywords=[3]{i1, i2, i4, i8, i16, i32, i64, half, float, double},
  morecomment=[l]{;},
  sensitive=true
}

\input{constants}
\input{pluverse-latex-style-guide/macro/pluverse-preamble-end.tex}

\AtBeginDocument{%
  \providecommand\BibTeX{{%
    \normalfont B\kern-0.5em{\scshape i\kern-0.25em b}\kern-0.8em\TeX}}}

\setcopyright{acmlicensed}
\copyrightyear{2018}
\acmYear{2018}
\acmDOI{XXXXXXX.XXXXXXX}

\acmConference[Conference acronym 'XX]{}{}{}
\acmISBN{978-1-4503-XXXX-X/18/06}

\pgfplotsset{compat=1.18}
\begin{document}

\title{Leveraging Large Language Models for Generalizing Peephole Optimizations}

\author{Chunhao Liao}
\orcid{0009-0004-4379-9454}
\affiliation{
   \institution{University of Waterloo}
     \country{Canada}
}
\email{chunhao.liao@uwaterloo.ca}

\author{Hongxu Xu}
\orcid{0009-0003-0926-220X}
\affiliation{
   \institution{University of Waterloo}
     \country{Canada}
}
\email{hongxu.xu@uwaterloo.ca}

\author{Xintong Zhou}
\orcid{0009-0002-6444-5431}
\affiliation{%
	\institution{University of Waterloo}
	\country{Canada}}
\email{x27zhou@uwaterloo.ca}

\author{Zhenyang Xu}
\orcid{0000-0002-9451-4031}
\affiliation{%
	\institution{University of Waterloo}
	\country{Canada}}
\email{zhenyang.xu@uwaterloo.ca}

\author{Chengnian Sun}
\orcid{0000-0002-0862-2491}
\affiliation{
   \institution{University of Waterloo}
   \country{Canada}
}
\email{cnsun@uwaterloo.ca}

\renewcommand{\shortauthors}{Liao, et al.}

\begin{abstract}
\input{abstract}
\end{abstract}



\maketitle

\section{Introduction}
\label{sec:intro}
\input{intro}

\section{Background}
\label{sec:background}

\input{background}

\section{Methodology}
\label{sec:methodology}

\input{methodology}

\section{Evaluation}
\label{sec:evaluation}
\input{evaluation}

\section{Related Work}
\label{sec:related-work}
\input{related-work}

\section{Conclusion}
\label{sec:conclusion}
\input{conclusion}

\bibliographystyle{ACM-Reference-Format}
\balance
\bibliography{acmart}

\newpage

\end{document}

%% file: pluverse-latex-style-guide/macro/pluverse-preamble.tex
\usepackage[ruled,lined,linesnumbered,vlined]{algorithm2e}

\SetCommentSty{mycommfont}
\DontPrintSemicolon 

\SetKwProg{Fn}{Function}{:}{}

\usepackage{amsmath}
\usepackage{amsfonts}
\usepackage{amsthm}
\usepackage{balance}
\usepackage{enumitem}
\usepackage{graphicx}
\usepackage{listings}
\usepackage{multicol}
\usepackage{multirow}
\usepackage{natbib}
\usepackage{subcaption}
\usepackage{url}
\usepackage{xspace}
\usepackage{xcolor}
\usepackage{tcolorbox}
\usepackage{colortbl}

\usepackage{tikz}
\usepackage{pgfplots}
\newcommand*\circled[1]{\tikz[baseline=(char.base)]{
    \node[shape=circle,draw,inner sep=0.5pt] (char) {\small#1};}}
\usetikzlibrary{shapes}
\usetikzlibrary{shapes.geometric}
\usetikzlibrary{arrows.meta, positioning}

\newcommand{\eg}{\mbox{\textit{e.g.}}\xspace}
\newcommand{\etal}{\mbox{\textit{et al.}}\xspace}
\newcommand{\etc}{\mbox{\textit{etc.}}\xspace}
\newcommand{\ie}{\mbox{\textit{i.e.}}\xspace}

\definecolor{BlindColorTolOne}{HTML}{332288}
\definecolor{BlindColorTolTwo}{HTML}{117733} 
\definecolor{BlindColorTolThree}{HTML}{44AA99}
\definecolor{BlindColorTolFour}{HTML}{88CCEE}
\definecolor{BlindColorTolFive}{HTML}{DDCC77}
\definecolor{BlindColorTolSix}{HTML}{CC6677} 
\definecolor{BlindColorTolSeven}{HTML}{AA4499}
\definecolor{BlindColorTolEight}{HTML}{882255}

\definecolor{BlindColorWongOne}{HTML}{000000} 
\definecolor{BlindColorWongTwo}{HTML}{E69F00}
\definecolor{BlindColorWongThree}{HTML}{56B4E9}
\definecolor{BlindColorWongFour}{HTML}{009E73}
\definecolor{BlindColorWongFive}{HTML}{F0E442}
\definecolor{BlindColorWongSix}{HTML}{0072B2} 
\definecolor{BlindColorWongSeven}{HTML}{D55E00}
\definecolor{BlindColorWongEight}{HTML}{CC79A7}

\definecolor{mygreen}{HTML}{02818a}

\mathchardef\mhyphen="2D

\newcounter{FindingCounter}
\newcommand{\finding}[1]{
	\begin{tcolorbox}
		\textbf{RQ\refstepcounter{FindingCounter}\theFindingCounter}: #1
	\end{tcolorbox}
}

\newcommand{\myparagraph}[1]{
  \vspace*{0.04cm}
  \noindent \textit{\textbf{#1.}}\quad
}

\newcommand{\projectname}[1]{\mbox{\textsf{#1}}\xspace} 
\SetKwData{AlgTrue}{true}
\SetKwData{AlgFalse}{false}
\SetKw{AlgBreak}{break}
\SetKw{AlgContinue}{continue}

%% file: constants.tex
\newcommand{\proj}{\projectname{LPG}}
\newcommand{\Hydradataset}{\textsc{HydraDataset}\xspace}
\newcommand{\projdataset}{\textsc{LPGDataset}\xspace}

\newcommand{\NumIntBenchmark}{41}
\newcommand{\NumFloatBenchmark}{16}
\newcommand{\NumMemBenchmark}{5}
\newcommand{\NumHydraBenchmark}{40}
\newcommand{\NumPeepGenHydraSuccessCount}{36}
\newcommand{\NumHydraHydraDatasetSuccessCount}{23}
\newcommand{\NumPeepGenPeepGenIntSuccessCount}{38}
\newcommand{\NumHydraPeepGenIntSuccessCount}{12}
\newcommand{\NumPeepGenFloatCount}{12}
\newcommand{\NumPeepGenMemCount}{4}
\newcommand{\NumPeepVsHydraBetterPeepGenDatasetIntCount}{8}
\newcommand{\NumPeepVsHydraWorsePeepGenDatasetIntCount}{1}
\newcommand{\NumPeepVsHydraBetterHydraDatasetCount}{14}
\newcommand{\NumPeepVsHydraWorseHydraDatasetCount}{1}

\newcommand{\ValIntBenchmark}{\NumIntBenchmark\xspace}
\newcommand{\ValFloatBenchmark}{\NumFloatBenchmark\xspace}
\newcommand{\ValMemBenchmark}{\NumMemBenchmark\xspace}
\newcommand{\ValprojBenchmark}{\number\numexpr\NumIntBenchmark+\NumFloatBenchmark+\NumMemBenchmark\relax\xspace}
\newcommand{\ValHydraBenchmark}{\NumHydraBenchmark\xspace}
\newcommand{\ValTotalBenchmark}{\number\numexpr\NumHydraBenchmark+\NumIntBenchmark+\NumFloatBenchmark+\NumMemBenchmark\relax\xspace}
\newcommand{\ValPeepVsHydraTotalInt}{\number\numexpr\NumHydraBenchmark+\NumIntBenchmark\relax\xspace}

\newcommand{\ValHydraHydraDatasetSuccessCount}{\NumHydraHydraDatasetSuccessCount\xspace}
\newcommand{\ValPeepGenHydraSuccessCount}{\NumPeepGenHydraSuccessCount\xspace}
\newcommand{\ValHydraPeepGenIntSuccessCount}{\NumHydraPeepGenIntSuccessCount\xspace}
\newcommand{\ValHydraIntCount}{\number\numexpr\NumHydraHydraDatasetSuccessCount+\NumHydraPeepGenIntSuccessCount\relax\xspace}
\newcommand{\ValPeepVsHydraBetterPeepGenDatasetIntCount}{\NumPeepVsHydraBetterPeepGenDatasetIntCount\xspace}
\newcommand{\ValPeepVsHydraWorsePeepGenDatasetIntCount}{\NumPeepVsHydraWorsePeepGenDatasetIntCount\xspace}
\newcommand{\ValPeepVsHydraBetterHydraDatasetCount}{\NumPeepVsHydraBetterHydraDatasetCount\xspace}
\newcommand{\ValPeepVsHydraWorseHydraDatasetCount}{\NumPeepVsHydraWorseHydraDatasetCount\xspace}
\newcommand{\ValPeepVsHydraBetterCount}{\number\numexpr\NumPeepVsHydraBetterHydraDatasetCount+\NumPeepVsHydraBetterPeepGenDatasetIntCount\relax\xspace}
\newcommand{\ValPeepVsHydraWorseCount}{\number\numexpr\NumPeepVsHydraWorseHydraDatasetCount+\NumPeepVsHydraWorsePeepGenDatasetIntCount\relax\xspace}

\newcommand{\ValPeepGenIntCount}{\number\numexpr\NumPeepGenHydraSuccessCount+\NumPeepGenPeepGenIntSuccessCount\relax\xspace}

\newcommand{\ValPeepGenFloatCount}{\NumPeepGenFloatCount\xspace}
\newcommand{\ValPeepGenMemCount}{\NumPeepGenMemCount\xspace}
\newcommand{\ValPeepGenOverallCount}{\number\numexpr\NumPeepGenHydraSuccessCount+\NumPeepGenPeepGenIntSuccessCount+\NumPeepGenFloatCount+\NumPeepGenMemCount\relax\xspace}

\newcommand{\ValPeepGenPeepGenIntSuccessCount}{\NumPeepGenPeepGenIntSuccessCount\xspace}
\newcommand{\ValCaseStudyNum}{3\xspace}

\newcommand{\ValPassConstSymEffective}{90\xspace}
\newcommand{\ValPassStructuralEffective}{23\xspace}

\newcommand{\ValPassRemovePCEffective}{13\xspace}
\newcommand{\ValPassWeakenPCEffective}{30\xspace}
\newcommand{\ValPassFlagAttributeRemovalEffective}{24\xspace}
\newcommand{\ValPassTypeGeneralizationEffective}{78\xspace}
\newcommand{\ValPassConstSymAffected}{216\xspace}
\newcommand{\ValPassStructuralAffected}{24\xspace}
\newcommand{\ValPassRemovePCAffected}{17\xspace}
\newcommand{\ValPassWeakenPCAffected}{34\xspace}
\newcommand{\ValPassFlagAttributeRemovalAffected}{48\xspace}

\newcommand{\ValRQThreeQwenTwoThirtyFiveInstructSuccessCount}{33\xspace}
\newcommand{\ValRQThreeQwenTwoThirtyFiveThinkingSuccessCount}{53\xspace}
\newcommand{\ValRQThreeQwenThirtyInstructSuccessCount}{28\xspace}
\newcommand{\ValRQThreeQwenThirtyThinkingSuccessCount}{37\xspace}

\newcommand{\mca}{llvm-mca\xspace}
\newcommand{\llc}{llc\xspace}
\newcommand{\alivetwo}{Alive2\xspace}

\newcommand{\smt}{SMT\xspace}
\newcommand{\llvm}{LLVM\xspace}
\newcommand{\ir}{IR\xspace}
\newcommand{\llm}{LLM\xspace}
\newcommand{\llms}{LLMs\xspace}
\newcommand{\instcombine}{InstCombine\xspace}
\newcommand{\hostArchitecture}{x86-64\xspace}

\newcommand{\hydra}{Hydra\xspace}
\newcommand{\minotaur}{Minotaur\xspace}

\newcommand{\zthree}{Z3\xspace}
\newcommand{\souper}{Souper\xspace}
\newcommand{\lampo}{Lampo\xspace}
\newcommand{\psyco}{PSyCO\xspace}
\newcommand{\aliveinfer}{Alive-Infer\xspace}
\newcommand{\pie}{PIE\xspace}
\newcommand{\pgen}{P-Gen\xspace}
\newcommand{\ungen}{Under-Generalization\xspace}

\newcommand{\gemini}{Gemini3-pro-preview\xspace}
\newcommand{\QBI}{Qwen3-235B-Instruct\xspace}
\newcommand{\QSI}{Qwen3-30B-Instruct\xspace}
\newcommand{\QBT}{Qwen3-235B-Thinking\xspace}
\newcommand{\QST}{Qwen3-30B-Thinking\xspace}

\newcommand{\instCombineCodeSize}{50,000\xspace}

\newcommand{\precondition}{\textit{PRE}\xspace}
\newcommand{\rhs}{\textit{RHS}\xspace}
\newcommand{\lhs}{\textit{LHS}\xspace}

\definecolor{BlindColorTolFive}{HTML}{DDCC77}

\definecolor{MotivConstColor}{HTML}{1F77B4}
\definecolor{MotivStructColor}{HTML}{ebb434}
\definecolor{MotivRemovePCColor}{HTML}{D62728}
\definecolor{MotivWeakenPCColor}{HTML}{9467BD}
\definecolor{MotivRemoveFlagsColor}{HTML}{2CA02C}
\definecolor{MotivBitwidthColor}{HTML}{27B7F5}
\newcommand{\MotivHighlightTint}{38}

\makeatletter
\newcommand{\mothighlight}[2]{%
  \begingroup
  \setlength{\fboxsep}{0.45pt}%
  \if\relax\detokenize{#2}\relax
    \def\mot@textcontent{\strut\hspace{1.2em}}%
    \def\mot@mathdisplay{\mathstrut\hphantom{\mathrm{xx}}}%
    \let\mot@mathtext\mot@mathdisplay
    \let\mot@mathscript\mot@mathdisplay
    \let\mot@mathscriptscript\mot@mathdisplay
  \else
    \def\mot@textcontent{#2}%
    \def\mot@mathdisplay{#2}%
    \let\mot@mathtext\mot@mathdisplay
    \let\mot@mathscript\mot@mathdisplay
    \let\mot@mathscriptscript\mot@mathdisplay
  \fi
  \ifmmode
    \mathchoice
      {\mathord{\colorbox{#1!\MotivHighlightTint}{\(\m@th\displaystyle \mot@mathdisplay\)}}}%
      {\mathord{\colorbox{#1!\MotivHighlightTint}{\(\m@th\textstyle \mot@mathtext\)}}}%
      {\mathord{\colorbox{#1!\MotivHighlightTint}{\(\m@th\scriptstyle \mot@mathscript\)}}}%
      {\mathord{\colorbox{#1!\MotivHighlightTint}{\(\m@th\scriptscriptstyle \mot@mathscriptscript\)}}}%
  \else
    \colorbox{#1!\MotivHighlightTint}{\mot@textcontent}%
  \fi
  \endgroup
}
\makeatother

\newcommand{\motconst}[1]{\mothighlight{MotivConstColor}{#1}}
\newcommand{\motstruct}[1]{\mothighlight{MotivStructColor}{#1}}
\newcommand{\motremovepc}[1]{\mothighlight{MotivRemovePCColor}{#1}}
\newcommand{\motweakenpc}[1]{\mothighlight{MotivWeakenPCColor}{#1}}
\newcommand{\motremoveflags}[1]{\mothighlight{MotivRemoveFlagsColor}{#1}}
\newcommand{\motbitwidth}[1]{\mothighlight{MotivBitwidthColor}{#1}}

\newcommand{\poweroftwo}{\text{PowerOfTwo}\xspace}
\newcommand{\popcount}{\text{PopCount}\xspace}
\newcommand{\bitwidth}{\text{Bitwidth}\xspace}

\newcommand{\leUnsigned}{\le_{\mathrm{u}}}
\newcommand{\leSigned}{\le_{\mathrm{s}}}
\newcommand{\ltSigned}{<_{\mathrm{s}}}
\newcommand{\ltUnsigned}{<_{\mathrm{u}}}
\newcommand{\geSigned}{\ge_{\mathrm{s}}}
\newcommand{\rightShiftUnsigned}{\gg_{\mathrm{u}}}
\newcommand{\zext}{\mathrm{zext}}
\newcommand{\sext}{\mathrm{sext}}

\newcommand{\nsw}{\mathrm{nsw}}
\newcommand{\smin}{\mathrm{smin}}
\newcommand{\smax}{\mathrm{smax}}
\newcommand{\cttz}{\mathrm{cttz}}
\newcommand{\fcmp}{\mathrm{fcmp}}
\newcommand{\oeq}{\mathrm{oeq}}
\newcommand{\andOp}{\mathbin{\&}}
\newcommand{\orOp}{\mathbin{|}}
\newcommand{\tyInt}[1]{\mathrm{i#1}}

\newcommand{\rqone}{Can \proj effectively generalize real-world peephole optimizations?}
\newcommand{\rqtwo}{What is the impact of each strategy on the generalization quality?}
\newcommand{\rqthree}{How do different large language models affect \proj's generalization performance?}

%% file: pluverse-latex-style-guide/macro/pluverse-preamble-end.tex
\usepackage{cleveref} 

\Crefname{algocf}{Algorithm}{Algorithms}
\crefname{algocf}{Algorithm}{Algorithms}

\Crefname{algorithm}{Algorithm}{Algorithms}
\crefname{algorithm}{Algorithm}{Algorithms}

\crefname{appendix}{Appendix}{Appendices}
\Crefname{appendix}{Appendix}{Appendices}

\Crefname{figure}{Figure}{Figures}
\crefname{figure}{Figure}{Figures}

\crefname{listing}{Listing}{Listings}
\Crefname{listing}{Listing}{Listings}

\Crefname{table}{Table}{Tables}
\crefname{table}{Table}{Tables}

\crefname{thm}{Theorem}{Theorems}
\Crefname{thm}{Theorem}{Theorems}

\crefname{equation}{Equation}{Equations}
\Crefname{equation}{Equation}{Equations}

\crefformat{chapter}{\S~#2#1#3}
\crefmultiformat{chapter}{\S\S~#2#1#3}{ and~#2#1#3}{, #2#1#3}{, and~#2#1#3}

\crefformat{section}{\S~#2#1#3}
\crefmultiformat{section}{\S\S~#2#1#3}{ and~#2#1#3}{, #2#1#3}{, and~#2#1#3}

%% file: abstract.tex
Peephole optimizations are a core component of modern optimizing compilers.
It rewrites specific instruction into semantically equivalent but more
efficient forms. In practice, creating a new peephole optimization often starts
from a concrete optimization instance and requires lifting it into a more
general rewrite rule that matches a wider range of instruction patterns. This
generalization step is critical to optimization effectiveness, but it is also
difficult: producing rules that are both correct and sufficiently general
typically demands substantial manual effort and domain expertise. Existing
approaches such as \hydra attempt to automate this task with program synthesis,
but their generalization capability is often limited by search-space explosion,
under-generalization, and restricted support for diverse instruction domains.

We present \proj, large language model aided peephole optimization generalization,
a framework that uses large language models (\llms) to
generalize peephole optimizations. The design of \proj is motivated by the
observation that \llms are effective at semantic abstraction and exploratory
reasoning, while formal analyses are necessary to ensure that generated rules
are sound and profitable. Based on this observation, \proj adopts a closed-loop
workflow that integrates \llm-driven symbolic constant generalization,
structural generalization, constraint relaxation, and bitwidth/precision
generalization with feedback from syntactic validation, semantic verification,
and profitability checking.

We evaluate \proj on real-world peephole optimization issues drawn from the
\llvm ecosystem. Overall, \proj successfully generalizes
\ValPeepGenOverallCount out of \ValTotalBenchmark optimizations. On the
integer-focused subset that is directly comparable to \hydra, \proj
generalizes \ValPeepGenIntCount out of \ValPeepVsHydraTotalInt optimizations,
whereas \hydra generalizes \ValHydraIntCount. Moreover, \proj produces
strictly more general rules in \ValPeepVsHydraBetterCount cases, compared with
\ValPeepVsHydraWorseCount for \hydra. These results show that \proj is more
effective at generalizing real-world peephole optimizations across a broad
range of instruction domains.

%% file: intro.tex
Peephole optimization is a cornerstone of modern compilation pipelines.
Operating on a sliding window of instructions (the ``peephole''),
this technique identifies and replaces specific instruction patterns
with semantically equivalent and more efficient alternatives~\cite{muchnick1997advanced, aho2006compilers, alive}.
Its significance is underscored by its substantial scale
in industrial compilers; for instance, \llvm dedicates over
\instCombineCodeSize lines of C++ code to peephole optimizations in its \instcombine pass alone~\cite{llvm_instcombine}.
However, previous studies highlight that peephole optimizers
are among the most error-prone components in a compiler~\cite{hydra, lpo, alive1, issta2016compilerbug}.
This is exemplified by \llvm's peephole optimizer,
which accounts for a significant share of reported miscompilations~\cite{yang2011finding, LLVM, emi}.
Consequently, developing effective and reliable peephole optimizers
remains both essential and challenging in modern compiler systems.

Conventionally, adding a new peephole optimization involves three steps~\cite{hydra}:
\circled{1}~identifying a suboptimal instruction sequence and deriving an optimized replacement;
\circled{2}~generalizing this specific transformation into a broadly applicable rewrite rule;
and \circled{3}~implementing the generalized rule within the compiler's codebase.
This generalization phase (\ie, step \circled{2}) is critical; it abstracts specific optimization instances
into rules that enable the compiler to handle diverse instruction sequences
beyond isolated concrete cases.

Compiler developers typically generalize optimizations
manually to extend their scope beyond initial motivating
examples~\cite{hydra, buchwald2015optgen, proofgen}.
However, this manual approach poses significant hurdles;
specifically, deriving general rules becomes increasingly difficult
as optimization logic grows more complex~\cite{llvm_issue_79690_comment,llvm_issue_generalization_139641,llvm_issue_86061,llvm_issue_88481,llvm_issue_97168,llvm_issue_12792,llvm_issue_51744}.
In one \llvm issue report regarding a missed optimization~\cite{llvm_issue_generalization_139641},
a maintainer explicitly admitted: {``\textit{I don't know how to generalize this specific fold.}''}
The fact that this issue remains unresolved long after being assigned
underscores the non-trivial nature of manual peephole generalization.
Furthermore, manual methods lack systematic rigor,
introducing risks of unsoundness.
This process is so error-prone
and demanding that valid optimizations are frequently abandoned
due to the prohibitive burden of manual generalization~\cite{hydra, buchwald2015optgen, proofgen}.

Prior work such as \hydra~\cite{hydra} has
sought to automate optimization generalization through
enumerative synthesis.
While these methods systematically explore candidate
generalizations, they suffer from three key limitations.
First, the combinatorial growth of the search space frequently leads to
synthesis failures (\eg, timeouts), especially for complex optimizations.
Second, the resulting optimizations are often under-generalized, \ie,
less general than what is theoretically attainable.
For example, \hydra adopts a greedy policy that returns the first valid
generalization it finds, which may not be the most general one in the search space.
Third, these techniques are typically restricted to narrow instruction
domains, primarily integer arithmetic, and struggle to generalize
optimizations involving floating-point or memory instructions.
A more detailed discussion of \hydra's limitations
is provided in \cref{subsec:peephole-optimization-generalizations}.

\myparagraph{\proj}
To bridge these gaps, we present \proj, an \llm-based pipeline for peephole
optimization generation.
\proj integrates
formal verification to ensure the correctness of generalized rewrite rules
and performance cost modeling to guarantee profitability~(\cref{subsec:verification}).
Specifially, \proj orchestrates generalization across four stages:
\emph{Symbolic Constant Generalization}~(\cref{subsec:symbolic-constant-generalization}),
\emph{Structural Generalization}~(\cref{subsec:structural-generalization}),
\emph{Constraint Relaxation} (\cref{subsec:constraint-relaxation}),
and \emph{Bitwidth/Precision Generalization}~(\cref{subsec:bitwidth-precision-generalization}).
By employing \llms throughout this multi-stage process,
\proj directly addresses the limitations of prior synthesis-based approaches.
First, \proj leverages the reasoning capabilities of \llms to infer program intent,
enabling it to explore large transformation spaces without
incurring search-space explosion.
Second, due to the semantic understanding provided by \llms,
\proj is not constrained by greedy policies and thus
has greater potential to derive more general rewrite rules.
Third, rather than restricting generalization
to integer-only peephole optimizations, \proj exploits the
generative flexibility of \llms to extend its scope to
more diverse and complex optimization domains, including
floating-point and memory instructions.

We extensively evaluated \proj's capability in generalizing
peephole optimizations within the \llvm ecosystem.
To facilitate a direct comparison with \hydra, we first evaluated
\proj on \Hydradataset, which consists of \ValHydraBenchmark benchmarks
used in the original \hydra study~\cite{hydra}. On \Hydradataset,
\proj successfully generalized \ValPeepGenHydraSuccessCount out
of \ValHydraBenchmark optimizations, whereas \hydra generalized
\ValHydraHydraDatasetSuccessCount. Moreover, among the cases on
which both systems succeeded, \proj produced strictly more
general rewrite rules in \ValPeepVsHydraBetterHydraDatasetCount
cases, while \hydra did so in only
\ValPeepVsHydraWorseHydraDatasetCount case.
To mitigate potential data leakage in evaluating \llm-based
approaches, we further curated \projdataset, which comprises
\ValprojBenchmark real-world missed optimizations mined from
\llvm GitHub issues reported after the knowledge cutoff dates of
the evaluated models.
On \projdataset,
\proj successfully generalized
\ValPeepGenPeepGenIntSuccessCount
out of \ValIntBenchmark integer optimizations,
\ValPeepGenFloatCount out of \ValFloatBenchmark
floating-point optimizations, and \ValPeepGenMemCount out of
\ValMemBenchmark memory optimizations,
while \hydra generalized only
\ValHydraPeepGenIntSuccessCount out of \ValIntBenchmark integer optimizations.
Among the cases where both systems succeeded,
\proj produced strictly more general rewrite rules in
\ValPeepVsHydraBetterPeepGenDatasetIntCount cases, whereas
\hydra did so in \ValPeepVsHydraWorsePeepGenDatasetIntCount
case.
Furthermore, we assessed the effectiveness of the proposed generalization
strategies within \proj. The results
underscore the critical role of these different strategies
in enhancing the \proj's generalization capabilities.

\noindent\textbf{\emph{Contributions.}} \quad
Our main contributions are listed below:
\begin{itemize}[leftmargin=*, topsep=0pt, partopsep=0pt]
    \item
    We introduced the use of \llms for peephole optimization generalization.
    By leveraging their semantic reasoning capabilities, our approach overcomes the
    inherent limitations of traditional program-synthesis-based methods.

    \item
    We designed and implemented \proj, a specialized pipeline for generalizing
    peephole optimizations. \proj integrates four novel generalization
    strategies with formal verification and feedback-guided refinement to
    ensure both the correctness and the broad applicability of the
    generated optimizations.

    \item We conducted a comprehensive evaluation of
    \proj using real-world optimizations in \llvm \ir ,
    demonstrating its ability to generalize optimizations
    across diverse instruction domains. In total, \proj
    successfully generalized \ValPeepGenOverallCount out of
    \ValTotalBenchmark cases. On the integer-related subset
    comparable with \hydra, \proj generalized
    \ValPeepGenIntCount out of \ValPeepVsHydraTotalInt cases,
    compared with \ValHydraIntCount for \hydra, and produced
    strictly more general rules in \ValPeepVsHydraBetterCount
    cases versus \ValPeepVsHydraWorseCount for \hydra.
    Furthermore, we investigated the effectiveness of the
    distinct generalization strategies in \proj's pipeline,
    and the results underscore their significant role in
    improving overall generalization effectiveness. Taken
    together, these results highlight the advantages of \proj
    over the state of the art and establish it as a broadly
    applicable, effective tool for compiler development.
\end{itemize}

%% file: background.tex
We built \proj upon \llvm to generalize peephole optimizations in \llvm
\ir~\cite{LLVM}.
As a modular, open-source compiler infrastructure,
\llvm utilizes a low-level, SSA-based, and platform-independent
\ir designed to support diverse source languages
and reusable toolchain components.
We selected \llvm due to its extensive tooling ecosystem
and its widespread adoption across both industry and academia.
While our current implementation specifically targets \llvm \ir,
the underlying principles of \proj
remain applicable to other compiler frameworks.
The following  provides
necessary background on peephole optimizations in \llvm,
their generalization, and  formal verification.

\subsection{Peephole Optimizations}
\label{subsec:peephole-optimizations-llvm}

Peephole optimization is a well-established compilation technique
for performing local code transformations.
By inspecting a small, sliding window of instructions,
the optimizer replaces specific sequences with semantically equivalent,
more efficient alternatives.
Formally, a peephole optimization is characterized
by the triple~\cite{muchnick1997advanced, cooper2011engineering, hydra}:
\[
\precondition \vDash \lhs \Rightarrow \rhs
\]
where \lhs (left-hand side) represents the target instruction sequence,
$\rhs$ (right-hand side) is its optimized replacement,
and $\precondition$ is a logical predicate over the operands in \lhs and \rhs
that defines the rewrite's validity.
Formally, This judgment states that substituting
$\lhs$ with $\rhs$ is semantics-preserving
provided that $\precondition$ holds.
If no explicit $\precondition$ is specified,
then $\precondition=\AlgTrue$, meaning that
the rewrite is unconditional
and there are no constraints
over the operands in \lhs and \rhs.

A classic peephole optimization example is strength reduction~\cite{aho2006compilers},
where multiplication by a power of two is replaced by a computationally cheaper left-shift:
\[
\poweroftwo(C) \vDash x \times C \Rightarrow x \ll \log_2(C)
\]
This substitution is preferred because bitwise shifts typically incur lower CPU latency and energy costs than multiplication and division.

In \llvm, peephole optimizations are mainly implemented within
\instcombine, a massive optimization pass written in C++ comprising
over \instCombineCodeSize lines of hand-written pattern-matching rules that identify and transform
specific instruction sequences~\cite{llvm_instcombine}.

\subsection{Peephole Optimization Generalization}
\label{subsec:peephole-optimization-generalizations}

Peephole optimization generalization is the systematic process of
abstracting a concrete optimization instance into a more general transformation rule that remains valid across a broader class of
programs.
While an initial optimization is often discovered from a specific code
fragment, the objective of generalization is to expand its applicability
domain without compromising semantic correctness.
A generalization is
considered successful if it preserves correctness on all cases
handled by the original optimization while also enabling
beneficial rewrites on additional cases outside the original
domain.
Formally, given two successful generalizations, one is considered more general if its applicability domain is a strict superset of the other’s.

To illustrate the concept of peephole optimization generalization,
consider the generalization for a concrete peephole optimization:
\[
\AlgTrue \vDash
((x \oplus 173) \andOp 94) \oplus 57
\Rightarrow
(x \andOp 94) \oplus 53
\]
where $\oplus$ denotes exclusive OR (XOR), $\andOp$ denotes
bitwise conjunction (AND).
This instance is difficult to generalize by human inspection.
Nevertheless, a generalization framework may be able to  infer the latent
relation among the constants and synthesize the generalized
rewrite
\[
C_4 = (C_1 \andOp C_2) \oplus C_3 \vDash
((x \oplus C_1) \andOp C_2) \oplus C_3 \Rightarrow
(x \andOp C_2) \oplus C_4
\]
This rule is derived by applying the distributive law of bitwise AND over XOR:
$
(A \oplus B) \andOp M = (A \andOp M) \oplus (B \andOp M).
$
Unlike the original concrete rewrite,
which is restricted to a single constant instance,
the generalized rule applies to all instantiations satisfying the precondition
 \(C_4 = (C_1 \andOp C_2) \oplus C_3\).
By strictly enlarging the set of programs to which the optimization applies, peephole
optimization generalization is critical to maximizing compiler performance across a
broad range of code patterns.

\myparagraph{Manual Generalization}
Traditional implementations require developers to manually
generalize each concrete peephole optimization.
However, hand-crafted process faces two challenges.
First, empirical analysis reveals that the manual generalization of
transformation rules constitutes a primary source of correctness
issues in peephole optimizations~\cite{hydra}.
Second, developers may not always succeed in deriving a sound
and sufficiently general rule from a specific optimization
instance, potentially leaving valid optimization opportunities
undiscovered.
To illustrate the second challenge, we revisit a missed
peephole optimization report from \llvm~\cite{llvm_issue_generalization_139641}
introduced in \cref{sec:intro},
where the maintainer explicitly remarks on
the difficulty of generalizing this fold.
The corresponding \lhs and \rhs \llvm \ir fragments are shown in~\cref{fig:llvm-missed-generalization}.
\begin{figure}[h]
\centering
\hspace*{0.08\linewidth}%
\begin{minipage}[t]{0.4\linewidth}
\begin{lstlisting}
define i64 @lhs(i32 %arg0) {
  %1 = sub i32 0, %arg0
  %2 = and i32 %1, 63
  %3 = zext nneg i32 %2 to i64
  %4 = sub nsw i64 0, %3
  %5 = lshr i64 %4, 8
  %6 = or i64 %5, %4
  ret i64 %6
}
\end{lstlisting}
\vspace{2pt}
{\small\textbf{(a) \lhs (pre-transformation)}}
\end{minipage}
\hspace{0.08\linewidth}%
\begin{minipage}[t]{0.4\linewidth}
\begin{lstlisting}
define i64 @rhs(i32 %arg0) {
  %1 = and i32 %arg0, 63
  %2 = icmp ne i32 %1, 0
  %3 = sext i1 %2 to i64
  ret i64 %3
}
\end{lstlisting}
\vspace{2.7\baselineskip}
\vspace{2pt}
{\small\textbf{(b) \rhs (post-transformation)}}
\end{minipage}
\caption{A missed peephole optimization example from \llvm.
In \llvm \ir, ``\texttt{sub}'' is subtraction instruction, ``\texttt{and}'' is bitwise AND, ``\texttt{zext}'' is zero-extension,
``\texttt{lshr}'' is logical right shift, ``\texttt{or}'' is bitwise OR, ``\texttt{icmp ne}'' is integer comparison for not-equal, and ``\texttt{sext}'' is sign-extension.
Flags like ``\texttt{nneg}'' and ``\texttt{nsw}'' indicate that the instruction does not allow certain behaviors such as negative overflow and signed overflow,
and may cause undefined behavior if they occur.
}
\label{fig:llvm-missed-generalization}
\end{figure}
Generalizing this peephole optimization is difficult mainly
because its validity depends on the interaction of several
components. Specifically, it requires simultaneous reasoning
about arithmetic negation, masking, cross-bitwidth normalization
(\texttt{i32} to \texttt{i64}), flag-sensitive behavior
(\texttt{nneg}, \texttt{nsw}) and \llvm poison semantics
within one reusable rule.
Notably, to assist \llvm developers in implementing this peephole
optimization, we posted \proj's generalized result of this peephole
optimization in the issue thread, and the \llvm maintainer
explicitly acknowledged its correctness and generality~\cite{llvm_issue_139641_comment_3979594887}.

\begin{table}[H]
  \centering
  \footnotesize
  \renewcommand{\arraystretch}{1.2}
  \begin{tabular}{
    @{}>{\raggedright\arraybackslash}p{0.05\linewidth}
    >{\arraybackslash}p{0.42\linewidth}
    >{\arraybackslash}p{0.44\linewidth}@{}
  }
    \toprule
    & \textbf{Original} & \textbf{Generalized} \\
    \midrule

    \precondition
    & $\AlgTrue$
    &
    $\begin{aligned}[t]
      &\poweroftwo(C_1 + 1) \\
      {}\land{}\;& 0 \leUnsigned V \leUnsigned C_1 \\
      {}\land{}\;& \popcount(C_1) \leUnsigned C_2 \leUnsigned W - \popcount(C_1)
    \end{aligned}$
    \\
    \midrule

    \addlinespace[0.4ex]
    \lhs
    &
    $\begin{aligned}[t]
      &\bigl((-\zext(-x)\andOp 63)\bigr)\rightShiftUnsigned 8
      \orOp -\zext(-x \andOp 63)
    \end{aligned}$
    &
    $(-V)\rightShiftUnsigned C_2 \orOp (-V)$
    \\\midrule

    \addlinespace[0.4ex]
    \rhs
    & $\sext(x \andOp 63 \neq 0)$
    & $\sext(V \neq 0)$
    \\
    \bottomrule

    \multicolumn{3}{@{}p{0.88\linewidth}@{}}{\footnotesize
    Note: $\popcount(C_1)$ denotes the number of 1-bits in $C_1$. \newline
    \texttt{sext} and \texttt{zext} are the sign-extend and zero-extend operations. \newline
    For operators with a subscript, such as $\rightShiftUnsigned$ and $\leUnsigned$,
    the subscript denotes the signedness of the operation (``u'' for unsigned, ``s'' for signed), which is relevant for overflow and comparison semantics in \llvm \ir.
    }
  \end{tabular}

  \caption{Original and generalized forms of the peephole optimization shown in~\cref{fig:llvm-missed-generalization}.}
  \label{fig:llvm-generalized-solution}
\end{table}

\myparagraph{Automated Generalization}
\hydra is the state-of-the-art technique
to automatically generalize peephole optimizations from
concrete optimization instances~\cite{hydra}.
Built upon the \souper superoptimizer~\cite{souper},
\hydra generalizes transformations through inductive and enumerative synthesis
guided by heuristics and greedy algorithms.
While it ensures soundness using \alivetwo~\cite{alive2},
several inherent limitations constrain its effectiveness:

\begin{itemize}[leftmargin=*, topsep=0pt, partopsep=0pt]

\item \textbf{Search-Space Induced Synthesis Failure}
\hydra relies on depth-bounded enumerative synthesis,
which is susceptible to exponential state-space explosions
as instruction sequences increase in complexity.
This often results in
synthesis failures (\eg timeouts)
before a valid generalized rule can be discovered.
Furthermore, these rigid depth bounds strictly limit \hydra's expressiveness:
complex preconditions requiring more than two conjunctions
typically lead to synthesis failure~\cite{hydra},
leaving more sophisticated optimization patterns out of reach.

\item \textbf{\ungen:}
\hydra relies on heuristic rules and greedy algorithms
to mechanically recombine syntactic fragments,
rather than deep semantic understanding of the code.
Consequently, it might fail to uncover a more
general form, yielding \emph{\ungen}.
For example, \hydra heuristically binds concrete constants to
variable bitwidths. When an optimization contains a constant
$c$ and a variable $v$, where $\bitwidth(v) = c$, \hydra greedily
infers that $c$ denotes
$\bitwidth(v)$ and returns a result even
though $c$ may be semantically unrelated to $v$'s width.
This creates an inherent dilemma:
On one hand, applying this heuristic rule may misbind the constant
and restrict the rule to a narrow domain where it holds only
when $\bitwidth(v) = c$, missing the truly general form; On the other hand,
disabling this heuristic rule may miss valid cases where the
constant genuinely denotes a variable's bitwidth.

\item \textbf{Limited Domain Support:}
\hydra is restricted to integer operations,
explicitly excluding floating-point and memory
instructions. This restriction stems from the prohibitive
complexity of the synthesis search space and verification challenges in these domains.
Consequently, \hydra misses a substantial spectrum of critical peephole
optimizations found in modern high-performance compilers.

\end{itemize}
Due to these limitations, for the illustrative example shown in~\cref{fig:llvm-missed-generalization},
\hydra does not produce a result.
Our observations of the \llvm community indicate
that manual generalization continues to be the dominant practice~\cite{llvm_issue_154242,llvm_issue_140639,llvm_issue_130088}.

\subsection{Peephole Optimization Verification}
\label{subsec:verification-tools}

Generalized peephole optimizations must be verified for
both semantic correctness and performance profitability.
There are various tools available for these verification tasks.

\myparagraph{\textbf{Formal Semantic Verification}}
\alivetwo is a bounded translation-validation tool for \llvm
\ir~\cite{alive2}. It is fully automatic through invoking
\zthree~\cite{z3} to solve the \smt~\cite{smt} queries for
refinement checking.
Given pre- and post-transformation \llvm \ir sequences, it checks
whether the post-transformation \ir sequence \emph{refines}
the pre-transformation \ir sequence: for every input
on which the pre-transformation \ir sequence is defined,
the post-transformation sequence must produce the same result
and must not introduce additional undefined behavior.

\myparagraph{\textbf{Performance Profitability Assessment}}
Semantic correctness does not guarantee performance improvement:
a transformation may cause performance regressions, such as
increased instruction latency or reduced throughput.
Accordingly, profitability assessment is typically required,
and there are primarily two approaches.
\circled{1} Heuristic-based cost model like
\hydra's~\cite{hydra} assigns static costs to instructions to estimate profitability.
However, static instruction costs are only a coarse proxy
for runtime behavior and may therefore misjudge profitability,
admitting performance-degrading generalizations.
\circled{2} Simulation-based performance analysis, \eg
\mca (Machine Code Analyzer)~\cite{llvm_mca} provided by \llvm,
offers a more accurate profitability assessment by
leveraging information from scheduling models in LLVM
to measure the performance of machine code in a specific CPU.
In practice, \llvm \ir is first
lowered to assembly with \llc~\cite{llc}, the \llvm compiler,
after which \mca analyzes the resulting code to obtain multiple performance
metrics, such as the number of micro-operations (uOps) and the
estimated cycle count.

%% file: methodology.tex
This section presents the methodology of \proj for synthesizing
generalized peephole optimizations from concrete optimization
instances. \proj operates as an automated, closed-loop
pipeline, and~\cref{fig:workflow} provides an overview of
the workflow.
\proj generalizes peephole optimizations through four stages:

\begin{figure}[t]
  \centering

  \includegraphics[width=0.85\linewidth]{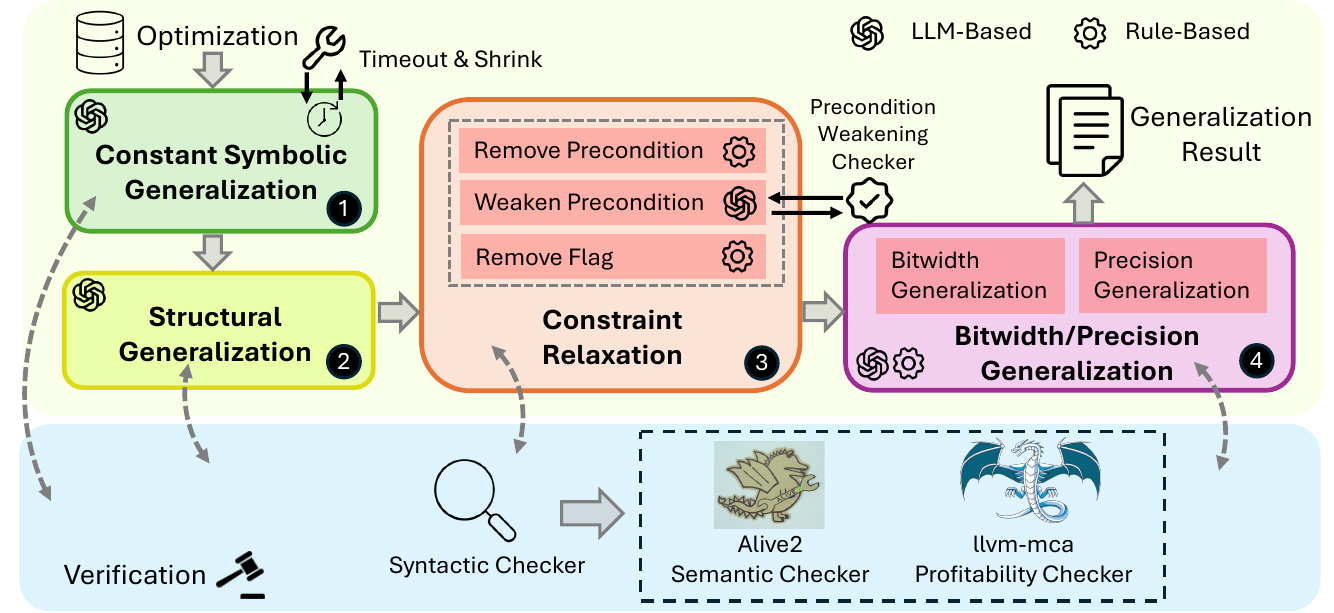}

  \caption{The overall workflow of \proj.
  }
  \label{fig:workflow}
\end{figure}

\begin{enumerate}[leftmargin=*, topsep=0pt, partopsep=0pt]
  \item \textbf{Constant Symbolic Generalization}~(\cref{subsec:symbolic-constant-generalization}).
        Replace concrete constants with symbolic constants.

  \item \textbf{Structural Generalization}~(\cref{subsec:structural-generalization}).
        Abstract fixed subexpressions (\eg $A+B$) with constrained symbolic values (\eg $V$),
        removing syntactic dependencies.

  \item \textbf{Constraint Relaxation}~(\cref{subsec:constraint-relaxation}).
        Enlarge
        the valid generalization domain by removing
        precondition, weakening precondition, and
        removing instruction flags and attributes.

  \item \textbf{Bitwidth/Precision Generalization}~(\cref{subsec:bitwidth-precision-generalization}).
        For integer optimizations, \proj performs
        bitwidth generalization, expanding a
        generalization in a specific bitwidth to a wider range of
        data types. For floating-point
        optimizations, \proj performs precision generalization,
        expanding a generalization in a specific precision to a wider
        range of floating-point types.
\end{enumerate}

\begin{table}[t]
  \centering
  \footnotesize

  \setlength{\tabcolsep}{2pt}
  \renewcommand{\arraystretch}{1.15}

  \begin{tabular}{@{}p{0.25\linewidth} p{0.55\linewidth} p{0.17\linewidth}@{}}

  \toprule

  \textbf{Step} & \textbf{Peephole Optimization} & \textbf{Transformation} \\

  \midrule

  Original Optimization~(\cref{subsec:peephole-optimizations-llvm}) &

  $\begin{aligned}
    & \left(0 -_{\nsw} \zext( -x : \tyInt{32} \andOp \motconst{63} ) \right) \rightShiftUnsigned \motconst{8} \orOp
       0 -_{\nsw} \zext( -x \andOp \motconst{63} ) \\
    \Rightarrow \; & \sext(x \andOp \motconst{63} \neq 0)
  \end{aligned}$ &

  \\
  \midrule

  \circled{1}
  Constant Symbolic \newline Generalization~(\cref{subsec:symbolic-constant-generalization}) &

  $\begin{aligned}
    & \motconst{C_1} : \tyInt{32} \geSigned 0 \land \poweroftwo(\motconst{C_1} + 1) \; \land \\
    & \popcount(\motconst{C_1}) \leUnsigned \motconst{C_2} : \tyInt{64} \leUnsigned 32 \\
    \vDash \; & (0 -_{\nsw} \motstruct{\zext(-x:\tyInt{32} \andOp \motconst{C_1})}) \rightShiftUnsigned \motconst{C_2} \orOp 0 -_{\nsw} \motstruct{\zext(-x \andOp \motconst{C_1})}\\
    \Rightarrow \; & \sext(\motstruct{-x\andOp\motconst{C_1}}\neq 0)
  \end{aligned}$ &

  $63 \to C_1$ \newline $8 \to C_2$

  \\
  \midrule

  \circled{2}
  Structural \newline Generalization~(\cref{subsec:structural-generalization}) &

  $\begin{aligned}
    & \motremovepc{C_1:\tyInt{32} \geSigned 0} \land \poweroftwo(C_1 + 1) \; \land\\
    & \popcount(C_1) \leUnsigned C_2:\tyInt{64} \leUnsigned 32 \land\\
    & 0 \leUnsigned \motstruct{V}:\tyInt{64} \leUnsigned \zext(C_1) \\
    \vDash \; & (0 -_{\nsw} \motstruct{V}) \rightShiftUnsigned C_2 \orOp 0-_{\nsw}\motstruct{V} \\
    \Rightarrow \; & \sext(\motstruct{V} \neq 0)
  \end{aligned}$ &

  $\zext(-x \andOp C_1) \to V$ \newline $(-x \andOp C_1) \to V$

  \\
  \midrule

\multicolumn{3}{l}{
\circled{3}
Constraint Relaxation~(\cref{subsec:constraint-relaxation})
}
  \\
  \addlinespace[0.4ex]

  \hspace{1em} Remove precondition
  &
  \hspace{0.5em}
  $\begin{aligned}
    & \poweroftwo(C_1 + 1) \; \land \\
    & \popcount(C_1) \leUnsigned C_2:\tyInt{64} \leUnsigned \motweakenpc{32} \; \land \\
    & 0 \leUnsigned V:\tyInt{64} \leUnsigned \zext(C_1) \\
    \vDash \; & (0 -_{\nsw} V) \rightShiftUnsigned C_2 \orOp 0 -_{\nsw} V
    \Rightarrow \sext(V \neq 0)
  \end{aligned}$ &

  remove $C_1 \geSigned 0$

  \\
  \cmidrule(l){2-3}
  \addlinespace[0.4ex]

  \hspace{1em} Weaken precondition
  &
  \hspace{0.5em}
  $\begin{aligned}
    & \poweroftwo(C_1 + 1) \; \land \\
    & \popcount(C_1) \leUnsigned C_2:\tyInt{64} \leUnsigned \motweakenpc{64 - \popcount(C_1)} \; \land \\
    & 0 \leUnsigned V:\tyInt{64} \leUnsigned \zext(C_1) \\
    \vDash \; & (0 \motremoveflags{-_{\nsw}} V) \rightShiftUnsigned C_2 \orOp 0 \motremoveflags{-_{\nsw}} V
    \Rightarrow \sext(V \neq 0)
  \end{aligned}$ &

  $32 \to$ \newline $64 -\popcount(C_1)$

  \\
  \cmidrule(l){2-3}
  \addlinespace[0.4ex]

  \hspace{1em} Flag \& Attribute Removal
  &
  \hspace{0.5em}
  $\begin{aligned}
    & \poweroftwo(C_1 + 1) \; \land \\
    & \popcount(C_1) \leUnsigned C_2 : \motbitwidth{\tyInt{64}} \leUnsigned \motbitwidth{64} - \popcount(C_1) \; \land \\
    & 0 \leUnsigned V:\motbitwidth{\tyInt{64}} \leUnsigned \zext(C_1) \\
    \vDash \; & (0 \motremoveflags{-} V) \rightShiftUnsigned C_2 \orOp (0 \motremoveflags{-} V)
    \Rightarrow \sext(V \neq 0)
  \end{aligned}$ &

  remove $\nsw$

  \\
  \midrule

  \circled{4}
  Bitwidth/Precision \newline Generalization~(\cref{subsec:bitwidth-precision-generalization}) &

  $\begin{aligned}
    & \poweroftwo(C_1 + 1) \; \land \\
    & \popcount(C_1) \leUnsigned C_2 \leUnsigned \motbitwidth{W} - \popcount(C_1) \; \land \\
    & 0 \leUnsigned V \leUnsigned C_1 \\
    \vDash \; & (-V) \rightShiftUnsigned C_2 \orOp (-V)
    \Rightarrow \sext(V \neq 0)
  \end{aligned}$ &

  $64 \to W$

  \\
  \midrule

  \addlinespace
  \multicolumn{3}{@{}p{0.96\linewidth}@{}}{\footnotesize Note: \motconst{} constant symbolic generalization; \motstruct{} structural generalization; \motremovepc{} remove precondition.} \\
  \multicolumn{3}{@{}p{0.96\linewidth}@{}}{\footnotesize \motweakenpc{} weaken precondition; \motremoveflags{} flag/attribute removal; \motbitwidth{} bitwidth/precision generalization.} \\
  \multicolumn{3}{@{}p{0.96\linewidth}@{}}{$W$ represents arbitrary \bitwidth; $\popcount(C)$ represents the number of 1-bits in $C$.} \\
  \bottomrule
  \end{tabular}
  \caption{
      Illustrating example: iterative generalization from the original peephole optimization instance.
  }
  \label{tab:motivating-generalization-pipeline}
\end{table}

\subsection{Illustrative Example}
We illustrate \proj's generalization process using the
peephole optimization in~\cref{fig:llvm-missed-generalization}.
\Cref{tab:motivating-generalization-pipeline} presents this
process step by step, where each step corresponds to one
generalization strategy
and finally derives the
generalization result in~\cref{fig:llvm-generalized-solution}.

\proj starts with a concrete optimization instance whose validity
depends on the literal constants $63$ and $8$, the specific
syntactic form $\zext(-x \andOp 63)$, the
\texttt{nsw} flags, and the fixed \llvm integer bitwidths.
At this point, the optimization is tied to this single concrete
example.
The objective of the workflow is to generalize this example
while preserving correctness.

\myparagraph{Step \circled{1}: Constant Symbolic Generalization}
In the constant symbolic generalization stage, \proj replaces the
concrete constants with symbolic constants $C_1$ and $C_2$ and
introduces preconditions that capture the properties
required by the optimization. The resulting candidate preserves
the original computation pattern, but it is no longer tied to the
single constant instance (63 or 8).

\myparagraph{Step \circled{2}: Structural Generalization}
\proj then performs structural generalization. Instead of
requiring the concrete subexpression
$\zext(-x \andOp C_1)$, it introduces a fresh
symbolic value $V$ together with the constraint
$0 \leUnsigned V \leUnsigned \zext(C_1)$.
This step removes dependence on a particular syntax tree and
retains only the semantic property needed by the optimization.

\myparagraph{Steps \circled{3}: Constraint Relaxation}
After obtaining this structurally generalized candidate, \proj
enters constraint relaxation. As shown in
\cref{tab:motivating-generalization-pipeline}, this stage proceeds
in three substeps. \proj first removes the redundant precondition
$C_1 \geSigned 0$. It then weakens the upper bound on
$C_2$ from the concrete constant $32$ to the more general
condition $64 - \popcount(C_1)$. Finally, it removes
the \texttt{nsw} flags. Each substep enlarges the applicability
domain by eliminating a restriction inherited from the original
optimization.

\myparagraph{Step \circled{4}: Bitwidth/Precision Generalization}
Finally, \proj performs bitwidth
generalization by replacing the fixed \llvm integer types with a
symbolic width parameter $W$. The resulting rule is no longer
tied to \texttt{i32} and \texttt{i64}; instead, it applies to
arbitrary integer bitwidths as long as the constraints among
$C_1$, $C_2$, and $V$ are satisfied. As reflected in the workflow
in~\cref{fig:workflow}, each intermediate candidate can then be
passed to the verification and downstream generation stages, and
the final result is the generalized peephole optimization
shown in~\cref{fig:llvm-generalized-solution}.

\subsection{Verification of Soundness and Profitability}
\label{subsec:verification}

In this paper, we incorporate verification throughout the
generalization pipeline, accepting a generalization candidate
only when it is both semantically sound and performance-profitable.
As illustrated in~\cref{fig:workflow}, before invoking the
downstream verification tools, \proj first applies
syntactic check to each generated candidate.
The malformed or structurally inconsistent candidates are
repaired through regeneration, guided by the syntactic error
messages from the checker.
Only syntactically well-formed candidates enter the subsequent
two-phase verification for semantic correctness and performance
profitability.

\myparagraph{\textbf{Formal Semantic Verification}}
To guarantee that each generalized peephole optimization is
semantically valid, we invoke \alivetwo~\cite{alive2}
introduced in \cref{subsec:verification-tools}
to check refinement correctness between the \lhs
and \rhs \llvm \ir sequence. We discard any
generalization candidate that violates semantic correctness
in \alivetwo.

\myparagraph{\textbf{Performance Profitability Assessment}}
As discussed in \cref{subsec:verification-tools}, heuristic
cost models provide only a coarse proxy for runtime behavior.
Therefore, after establishing semantic correctness with
\alivetwo, \proj assesses profitability using \mca~\cite{llvm_mca}.
We lower \llvm \ir to assembly for the host architecture via
\llc~\cite{llc} and then invoke \mca to analyze the generated code.
We apply \hostArchitecture architecture in our experiment as
it is widely adopted in both industrial practice and academic
research.
Following prior \llvm superoptimization work~\cite{Minotaur},
we use uOps as the primary cost metric.
Consequently, we retain a transformation only when it is
profitable: the \rhs code sequence exhibits
a lower uOps count than the \lhs code sequence.

\myparagraph{\textbf{Verification Feedback Loop}}
After generating candidate generalizations, \proj runs
an automated refinement loop driven by verification outcomes.
For candidates that fail the verification, the resulting diagnostics
(\eg syntax errors, \alivetwo counterexamples, and \mca
regression reports) are fed back to the \llm to refine
subsequent generalizations.
Currently, we only include the feedback loop in the
first stage, Symbolic Constant Generalization,
mainly because it serves as the foundation for subsequent stages
and is more amenable to iterative refinement.

\subsection{Symbolic Constant Generalization}
\label{subsec:symbolic-constant-generalization}

Symbolic constant generalization abstracts concrete constants within a
peephole optimization into symbolic expressions,
broadening the applicability of the optimization beyond specific
literal values.
\proj utilizes \llm to propose these candidate generalizations.
Leveraging its semantic reasoning capabilities,
the \llm infers both the structural role of the constant and
the necessary precondition for validity, such as bitwidth,
value-range, or shift-legality constraints.
For example, a concrete shift amount $C$ can be abstracted to a symbolic
variable bounded by the operation's bitwidth $W$.
In contrast, \hydra relies on heuristic rules
rules and greedy algorithms within a bounded search space,
inherently limiting its generalization scope.
Moreover, \hydra conservatively avoids symbolizing certain constants
(\eg $0$, $1$, and $-1$), because abstracting them without deep
semantic context often yields invalid candidates.
\proj overcomes this limitation, using the \llm to safely
generalize these special constants when appropriate.

\myparagraph{\textbf{Recursive Bitwidth/Precision Reduction}}
For particularly complex candidates, soundness verification
with \alivetwo may time out.
\alivetwo performs refinement checking by encoding program
logic into \smt queries over bit-vector semantics.
Such queries may become computationally intractable due to
the explosion of the state space, which scales significantly
with the increase in integer bitwidth or floating-point precision.
To mitigate the risk that valid
generalizations are rejected due to solver resource limits,
\proj applies recursive bitwidth/precision reduction:
integer widths are halved (\eg \texttt{i16} $\to$
\texttt{i8}), floating-point types are progressively demoted
(\eg \texttt{double} $\to$ \texttt{float} $\to$
\texttt{half}). The 1-bit integer type is excluded from this
process because it typically represents boolean predicates (true/false).
Each reduction step uniformly scales bitwidths by the same
factor (i.e., half), thereby preserving their relative
proportionality.
After each reduction step, \proj re-runs verification to validate
the generalization at the reduced \bitwidth until it no longer
times out or a reduction limit is reached (\eg a literal can
no longer be encoded at the reduced width, or an odd bitwidth
prevents further halving \etc).
If verification fails or a reduction limit is reached,
but \alivetwo still times out, the generalization attempt is deemed unsuccessful.

\subsection{Structural Generalization}
\label{subsec:structural-generalization}

Structural generalization abstracts the fixed expression
structure of an optimization by replacing intermediate
computations with symbolic inputs, removing
dependencies on a particular syntax tree while retaining the
precondition necessary for correctness.
\Cref{fig:structural-generalization-mask-example} illustrates a
simple example. Structural generalization replaces the
subexpression $x \andOp 127$ with a symbolic value
$V$ and makes its semantic effect explicit through the range
constraint $0 \leUnsigned V \leUnsigned 127$. This
constraint implies that $V$ is non-negative, so
$0 - V \ltSigned 0$ is equivalent to $V \neq 0$: the predicate
is false when $V = 0$ and true for every $1 \leUnsigned V \leUnsigned 127$.
Hence the optimization
$\sext(0 - V \ltSigned 0) \Rightarrow \sext(V \neq 0)$ is valid. The precondition is essential
since the optimization fails for negative values of $V$.

\begin{table}[H]
\centering
\footnotesize
\renewcommand{\arraystretch}{1.2}
\begin{tabular}{l c c}

\toprule

& \textbf{Original} & \textbf{After Structural Generalization} \\

\midrule

\precondition &
\centering \AlgTrue &
$0 \leUnsigned \motstruct{V}:\tyInt{8} \leUnsigned 127$\\


\lhs &
$\sext(0-\motstruct{x:\tyInt{8} \andOp 127} \ltSigned 0)$ &
$\sext(0-\motstruct{V} \ltSigned 0)$\\


\rhs &
$\sext(\motstruct{x:\tyInt{8} \andOp 127} \neq 0)$ &
$\sext(\motstruct{V} \neq 0)$\\

\bottomrule

\end{tabular}
\caption{Structural generalization replaces the preserved
subexpression $x \andOp 127$ with a symbolic value
$V$ constrained by the range implied by the mask.}
\label{fig:structural-generalization-mask-example}
\end{table}

The key challenge lies in identifying subexpressions amenable
to abstraction and preserving the implicit constraints they
encode, so that the generalized optimization remains valid.
\proj prompts an \llm to conduct structural generalization.
Specifically, \proj guides the \llm to decouple instruction
structure from data dependencies with a task-specific prompt.
\proj directs the model to abstract the subexpressions
guarded with either \emph{non-relational precondition} or
\emph{relational precondition}.
Non-relational precondition includes
property abstractions such as KnownBits (the known zero and one bits of a value),
IntegerRange (the range of possible values for a variable), \etc or a collection of
facts such as \poweroftwo and $\textrm{LowBitsZero}_k$ (the k low bits of a value that are known to be zero).
For relational precondition, \proj instructs the model to
synthesize necessary precondition between values (\eg $x<y$).
In contrast, \hydra restricts precondition to a small set of
fix and predefined \emph{non-relational} preconditions,
which limits its expressiveness when an optimization's implicit
constraints cannot be captured by per-value properties.

\subsection{Constraint Relaxation}
\label{subsec:constraint-relaxation}

In this stage, \proj enlarges the applicability of an optimization by relaxing
the conditions under which it can be applied.
Specifically, \proj considers three kinds of generalization:
removing preconditions, weakening preconditions, and removing instruction flags
and attributes.
Each candidate generalization is validated as described
in \cref{subsec:verification} to ensure correctness.

This stage proceeds in three substeps:

\begin{itemize}[leftmargin=*]

    \item \textbf{Remove Precondition:}
    \proj enumerates the existing preconditions and attempts to remove them
    one at a time.
    A removal is retained only if the resulting optimization passes verification.

    \item \textbf{Weaken Precondition:}
    For preconditions that cannot be removed entirely, \proj leverages the
    \llm to propose weaker alternatives that admit more inputs.
    For example, a constraint $C > 0$ may be relaxed to $C \ge 0$.
    A weakened precondition is retained only if it passes verification and is
    confirmed to be \emph{strictly weaker} than the original one.

    \item \textbf{Remove Flags and Attributes:}
    \proj also attempts to remove instruction-level flags and related
    attributes that may unnecessarily restrict the optimization, including
    integer overflow flags (\eg \texttt{nsw}),
    floating-point fast-math flags (\eg \texttt{fast}),
    and function attributes (\eg \texttt{readonly}).
    As before, a removal is retained only if it passes verification.

\end{itemize}

Among these substeps, weakening preconditions is particularly challenging.
Prior enumerative approaches (\eg \hydra) weaken preconditions only within
a limited set of dataflow domains, such as \texttt{KnownBits}.
While it might be effective for some simple cases,
it is often inadequate for real-world optimizations,
whose preconditions frequently depend on richer semantic
relationships that are not expressible in these abstractions.
More generally, the space of useful relaxed predicates is
large and often continuous, making it difficult to enumerate
efficiently. For instance, relaxing an integer range requires
deciding not only whether a bound should be removed, but also
how far it can be widened. \proj addresses this challenge by
leveraging the \llm's semantic reasoning to propose candidate
relaxations in this broader search space.

\myparagraph{Strictly Weakened Generalization}
For precondition weakening, semantic preservation alone is insufficient:
the new precondition must also be strictly weaker than the original one.
To check this property, \proj uses a bidirectional refinement test in
\alivetwo.
In \alivetwo, a \precondition is encoded using additional \texttt{assume}
intrinsics in the \llvm \ir.
If an input does not satisfy the \precondition, the corresponding
\texttt{assume} is violated, causing the program to exhibit
\emph{undefined behavior} (UB).
Since a UB-triggering program is semantically equivalent to any program, it can
be refined by any other program.
Suppose the original precondition is $P$ and the weakened precondition is $P'$,
where $P \Rightarrow P'$.
Then the \lhs program under $P$ can be refined by the \lhs program under $P'$,
because every input satisfying $P$ also satisfies $P'$.
However, if the weakening is strict, the converse refinement does not hold:
there exist inputs satisfying $P'$ but not $P$.
Therefore, \proj checks refinement in both directions and accepts the candidate
only when the original version refines the weakened one, but not vice versa.
This ensures that the new precondition is indeed strictly weaker.

\subsection{Bitwidth/Precision Generalization}
\label{subsec:bitwidth-precision-generalization}

In this stage, \proj generalizes optimizations across different integer
bitwidths and floating-point precisions. This is crucial for
maximizing the applicability of optimizations, as many optimizations
are valid across a range of bitwidths or precisions, and tying them to
a specific type instance limits their utility.

However, proving an optimization at a single integer bitwidth or a single
floating-point precision does not imply that it is valid for all
type variants. In \llvm \ir, both integer and floating-point types
must be given as explicit concrete types (\eg \texttt{i8},
\texttt{i16}, \texttt{float}, \texttt{double});
symbolic forms such as \texttt{iN} or abstract precision variables
are not valid \llvm \ir syntax. Moreover, the verification toolchains
also operate only on concrete type instances.
Therefore, we cannot directly verify a generalized optimization over
arbitrary bitwidths or precision.

\myparagraph{Bitwidth Generalization}
For integer-related optimizations, \proj performs
bitwidth generalization using similar workflow as \hydra.
First, \proj runs a conservative static analysis, where
candidates containing explicit width-changing
casts (\texttt{trunc}, \texttt{sext}, \texttt{zext}) are rejected, and
only the trivial literals $\{0, 1, -1\}$ are allowed.
Second, we erase all concrete integer types in the candidates,
and then instantiate them with every concrete bitwidth,
up to a configurable maximum width (currently set to 64 bits),
so that we can verify every resulting instantiation individually.
Third, if brute-force instantiation succeeds only for a strict
subset of bitwidths, \proj uses \llm to synthesize an
additional predicate over the bitwidths (\eg $W_1 \le 8, W_1/W_2=2$).
\proj accepts the generalization only when
all bitwidth instantiations under the predicate pass the verification.

\myparagraph{Precision Generalization}
For floating-point optimizations, \proj performs
precision generalization: for a candidate rule
parameterized by precision $P$, we instantiate $P$ with
a predefined set of floating-point types.
We directly brute-force all three type instances:
\texttt{half}, \texttt{float}, and \texttt{double}.

%% file: evaluation.tex
This section presents our extensive evaluation of \proj,
demonstrating its effectiveness in generalizing real-world peephole optimizations.
We organize our evaluation
around three key research questions:

\begin{itemize}[leftmargin=*]
    \item
    \textbf{RQ1:} \rqone
    \item \textbf{RQ2:} \rqtwo
    \item \textbf{RQ3:} \rqthree
\end{itemize}

\subsection{Experimental Setup}

In this section, we describe the experimental setup for evaluating \proj,
including the hardware settings, selected models, data collection
and preprocessing procedures, and the evaluation metrics.

\subsubsection{Hareware Setup}
All experiments were conducted on a Ubuntu 22.04 (64-bit) server,
which is equipped with an Intel Xeon Gold 5217 CPU@3.00GHz with 32 cores,
and 376 GB of RAM.

\subsubsection{Selected Models}
\label{subsec:selected-models}
Table~\ref{tab:selected-models} summarizes the models used in our
evaluation. In RQ1 and RQ2, we use \gemini as the default model,
so that the effectiveness evaluation and strategy study are conducted
under a fixed model configuration. In RQ3, we evaluate four different
models: \QBI, \QBT,
\QST, and \QSI, to examine
how model scale and reasoning capability may affect \proj's generalization
performance. \gemini has a knowledge cutoff date of January 2025,
whereas all four Qwen3 variants have knowledge cutoff dates of
April 2025.

\begin{table}[ht]
\centering
\footnotesize
\setlength{\tabcolsep}{5pt}
\renewcommand{\arraystretch}{1.1}
\begin{tabular}{p{0.4\linewidth}cc}
\toprule
\textbf{Model} & \textbf{Reasoning} & \textbf{Knowledge Cutoff Date} \\
\midrule
\gemini & Yes & 01/2025 \\
\QSI & No & 04/2025 \\
\QST & Yes & 04/2025 \\
\QBI & No & 04/2025 \\
\QBT & Yes & 04/2025 \\
\bottomrule
\end{tabular}
\caption{Selected models for evaluation}
\label{tab:selected-models}
\end{table}

\subsubsection{Data Collection}
\label{data-collection}
To construct a robust test suite for assessing \proj's effectiveness,
we curated two datasets, namely \Hydradataset and \projdataset,
consisting of \ValTotalBenchmark distinct
peephole optimization instances collected from the \llvm project's GitHub issue tracker.
To ensure the quality in the test suite, each peephole
optimization instance underwent rigorous verification: we
filtered out invalid cases and retained only those that passed
the verification of soundness and profitability (\cref{subsec:verification}).

\vspace*{0.04cm}
  \noindent \textbf{\Hydradataset}\quad
For a direct comparison with \hydra~\cite{hydra},
\Hydradataset replicates all the instances used in the evaluation
from the original \hydra paper, comprising \ValHydraBenchmark
integer-related peephole optimizations collected from the \llvm GitHub issues.

\vspace*{0.04cm}
  \noindent \textbf{\projdataset}\quad
To validate \proj's effectiveness across diverse optimization domains
and to eliminate the risk of data leakage,
we curated an additional dataset,
\projdataset, by mining \ValprojBenchmark new peephole
optimizations from the \llvm GitHub issues. \projdataset
covers optimizations across multiple instruction categories,
including \ValIntBenchmark integer, \ValFloatBenchmark
floating-point, and \ValMemBenchmark memory optimizations.
Importantly, all instances in \projdataset were collected from
issues reported after May 2025, which is later than the
knowledge cutoff dates of the evaluated \llms.
This ensures that \proj's performance is assessed based on its
reasoning capabilities rather than memorization of training data.

\subsubsection{Data Preprocessing}
\label{subsec:preprocessing}

\begin{figure}[h]
\centering
\begin{subfigure}[t]{0.4\linewidth}
\centering
\begin{lstlisting}
define i8 @lhs(i64 %arg0) {
  %v0 = trunc i64 %arg0 to i8
  %v1 = urem i8 %v0, 25
  %v2 = urem i8 %v1, 5
  ret i8 %v2
}

define i8 @rhs(i64 %arg0) {
  %v0 = trunc i64 %arg0 to i8
  %v2 = urem i8 %v0, 5
  ret i8 %v2
}
\end{lstlisting}
\vspace{2pt}
\caption{Peephole Optimization Before Pruning}
\label{fig:pruning-example-before}
\end{subfigure}
\hspace{0.06\linewidth}
\begin{subfigure}[t]{0.4\linewidth}
\centering
\begin{lstlisting}
define i8 @lhs(i8 %newvar_v0) {
  %v1 = urem i8 %newvar_v0, 25
  %v2 = urem i8 %v1, 5
  ret i8 %v2
}

define i8 @rhs(i8 %newvar_v0) {
  %v2 = urem i8 %newvar_v0, 5
  ret i8 %v2
}
\end{lstlisting}
\vspace{1.8\baselineskip}
\vspace{2pt}
\caption{Peephole Optimization After Pruning}
\label{fig:pruning-example-after}
\end{subfigure}
\caption{An example of pruning for the peephole optimization.
In \llvm \ir, \texttt{trunc} is the truncation instruction that converts a value to a smaller type,
and \texttt{urem} is the unsigned remainder instruction.
}
\label{fig:pruning-example}
\end{figure}

This step aims to remove the noise in raw real-world optimization
instances, which frequently contain code irrelevant to the
optimization.
To isolate the core optimization logic, we adopt the
verification-guided pruning procedure of \hydra~\cite{hydra}.
Following this idea, we apply simplification to remove
unnecessary data-flow dependencies.
For each optimization instance, represented as a pair of functions
(\lhs, \rhs), we perform repeated pruning iteration over both
functions.
In each iteration, we consider one variable definition at a
time and tentatively replace all uses of that variable with
a fresh variable introduced as an additional function argument.
\cref{fig:pruning-example} provides an concrete example of the pruning process,
where \cref{sub@fig:pruning-example-before} and \cref{sub@fig:pruning-example-after}
show the optimization instance before and after pruning, respectively.
Pruning replaces an
intermediate value computed in the original instance
(\texttt{\%v0}) with a fresh argument (\texttt{\%newvar\_v0}).
During pruning, each tentative replacement is verified for soundness and profitability
and accepted only if both verification stages succeed; otherwise, it is reverted.
After each accepted substitution, dead instructions are eliminated,
and the pruning procedure repeats until an iteration introduces
no further changes.
Thus, every retained instruction is necessary to preserve semantic
correctness and performance profitability.

\subsection{RQ1: \rqone}
\label{subsec:rq1}
To quantitatively evaluate \proj's generalization capability
on real-world peephole optimizations, we conducted experiments
on all optimization instances from \Hydradataset and \projdataset
described in~\cref{data-collection}.
To evaluate \proj against the state-of-the-art,
we compared it with \hydra, a prior generalization approach restricted to integer optimizations.
We executed \hydra's publicly available implementation~\cite{hydra_repo}
on both its original dataset (\Hydradataset) and
the integer-related peephole optimizations from \projdataset.
We provide each tool with the concrete peephole
optimization instance and task it with generating a
generalized peephole optimization.

\subsubsection{Results}

\begin{table}[t]
\centering
\footnotesize
\caption{Peephole optimization generalization results of
\proj across instruction domains.
}
\label{tab:rq1-overall}
\begin{tabular}{lrr}
\toprule
\textbf{Instruction Domain} & \textbf{\#Instances} & \textbf{\#Success} \\
\midrule
Integer Optimization & \ValPeepVsHydraTotalInt & \ValPeepGenIntCount \\
Floating-point Optimization & \ValFloatBenchmark & \ValPeepGenFloatCount \\
Memory Optimization & \ValMemBenchmark & \ValPeepGenMemCount \\
\midrule
Overall & \ValTotalBenchmark & \ValPeepGenOverallCount \\
\bottomrule
\end{tabular}
\end{table}

\cref{tab:rq1-overall} summarizes the overall effectiveness of \proj
across all instances in our datasets, broken down by instruction domain.
In general,
\proj successfully synthesized \ValPeepGenOverallCount
generalized rewrite rules that pass the verification
out of the all \ValTotalBenchmark instances.
More specifically, \proj succeeded on
\ValPeepGenIntCount out of \ValPeepVsHydraTotalInt integer
optimizations, \ValPeepGenFloatCount out of \ValFloatBenchmark
floating-point optimizations, and \ValPeepGenMemCount out of
\ValMemBenchmark memory optimizations. These results suggest that
\proj generalizes beyond the integer-centric setting addressed by prior
work and can synthesize correct and profitable rules across diverse
instruction domains.

\begin{table}[t]
\centering
\caption{Comparison of \proj and \hydra on integer-related peephole optimizations.
}
\label{tab:rq1-proj-vs-hydra}
\footnotesize
\begin{tabular}{lrrrrr}
\toprule
\textbf{Dataset} & \textbf{\#Inst.} & \shortstack[r]{\#\textbf{\proj}\\ \textbf{Succ.}} & \shortstack[r]{\textbf{\#\hydra}\\ \textbf{Succ.}} & \shortstack[r]{\textbf{\#\proj} \\\textbf{Better}} & \shortstack[r]{\textbf{\#\hydra} \\ \textbf{Better}} \\
\midrule
\Hydradataset & \ValHydraBenchmark & \ValPeepGenHydraSuccessCount & \ValHydraHydraDatasetSuccessCount & \ValPeepVsHydraBetterHydraDatasetCount & \ValPeepVsHydraWorseHydraDatasetCount \\
\projdataset (Integer) & \ValIntBenchmark & \ValPeepGenPeepGenIntSuccessCount & \ValHydraPeepGenIntSuccessCount & \ValPeepVsHydraBetterPeepGenDatasetIntCount & \ValPeepVsHydraWorsePeepGenDatasetIntCount \\
\midrule
Overall & \ValPeepVsHydraTotalInt & \ValPeepGenIntCount & \ValHydraIntCount & \ValPeepVsHydraBetterCount & \ValPeepVsHydraWorseCount \\
\bottomrule
\end{tabular}
\end{table}

\Cref{tab:rq1-proj-vs-hydra} shows the comparison between
\proj and \hydra on integer-related peephole optimizations.
\proj successfully generalizes \ValPeepGenIntCount out of
\ValPeepVsHydraTotalInt instances
(\ValPeepGenHydraSuccessCount out of \ValHydraBenchmark on \Hydradataset
and \ValPeepGenPeepGenIntSuccessCount out of \ValIntBenchmark on \projdataset),
whereas \hydra succeeds on only \ValHydraIntCount instances
(\ValHydraHydraDatasetSuccessCount on \Hydradataset
and \ValHydraPeepGenIntSuccessCount on \projdataset).
Moreover, every instance successfully generalized by
\hydra is also successfully generalized by \proj. Restricting
to the cases both tools successfully generalizes,
\proj produces \emph{strictly more general} rewrite rules
in \ValPeepVsHydraBetterCount cases,
while \hydra does so in only \ValPeepVsHydraWorseCount cases.

\subsubsection{Analysis}
Beyond overall effectiveness, the two approaches show
distinctly different patterns in failure scenarios.
In particular, \proj exhibits a relatively concentrated error
profile: most failures are confined to a small number of
semantically difficult cases, especially those involving subtle
floating-point reasoning or limited verifier support.
By contrast, \hydra shows a broader range of failures arising
from different stages of its synthesis pipeline. This distinction
suggests that \proj is mainly limited by hard corner cases,
whereas \hydra is more often constrained by general search and
synthesis bottlenecks.

\myparagraph{\proj}
\proj successfully generalized the large majority of instances
across the evaluated benchmarks, indicating that its
generalization pipeline is robust on most real-world
optimizations.
The remaining non-successful cases are concentrated in only a few
semantically difficult optimizations.
More specifically, these failures comprise a small number of
semantic counterexamples together with one verification-inconclusive
case.
The counterexample cases indicate that \proj is most challenged
when a correct generalization depends on subtle IEEE-754 semantics
that are difficult to infer from the local rewrite alone, including
signed-zero distinctions, NaN-sensitive comparison and
\texttt{maxnum}-style behavior, and threshold rewrites whose
correctness depends on avoiding overflow, underflow, and rounding
error in derived floating-point constants~\cite{llvm_issue_153991,llvm_issue_85267,llvm_issue_186300}.
These optimizations may also depend on subnormal behavior and the
active rounding mode, so algebraically plausible rewrites need not
remain exact under floating-point semantics.
For the verification-inconclusive case, the difficulty does not
appear to stem from an obviously incorrect generalization, but
rather from incomplete verifier support for floating-point intrinsic
combinations such as \texttt{frexp} and \texttt{ldexp}~\cite{llvm_issue_186554}.
Taken together, these observations suggest that \proj is
limited primarily by a very small set of
optimizations at the boundary of floating-point reasoning,
memory semantics, and current verifier support.

\myparagraph{\hydra}
By contrast, although \hydra also succeeded on a
fraction of the benchmark suite, its behavior is noticeably less
stable. Its failures arise from several different stages in the
generalization pipeline, including repeated shrinking failures,
constant-synthesis exhaustion, timeouts, and crashes.
More importantly, these failures are not confined to a small set of
semantically difficult rewrites. Instead, they occur across a much
broader range of rewrite patterns and program contexts.
For example, on one optimization~\cite{llvm_issue_59519},
\hydra fails to produce a usable rule because the synthesized candidate
is ultimately rejected as ``not valid for any width.''
Even when \hydra does return a candidate, the result is often still
too specific, rather than capturing the underlying optimization logic.

\subsubsection{Case Studies}

In this section, we present \ValCaseStudyNum representative
case studies comparing \proj and \hydra, analyzing \ValCaseStudyNum
key dimensions of their generalization capabilities.
Each optimization in these case studies is extracted
from real-world optimization reports in the \llvm GitHub issues.

\myparagraph{Case Study 1: Search-Space Induced Synthesis Failure}
\cref{tab:case-study-timeout-search-space} presents a representative
instance~\cite{llvm_issue_157315} where \hydra timed out due to search-space explosion.
Concretely, \hydra exhausted its constant-synthesis budget in the synthesis loop.
This happens because a correct generalization must be expressed
as a parameterized rewrite: the pre-optimized instruction sequence
performs a clamp-based range check via \texttt{smax}/\texttt{smin}, while
the equivalent post-optimized instruction sequence is an unsigned range
test of the form $x+(C_1-C_2) \ltUnsigned C_3-C_2+1$ under semantic
precondition (e.g., $C_2 \le C_3$ and $(C_3-C_2) \neq -1$). The rewrite
also introduces affine constraints that couple multiple constants,
which make it difficult for \hydra to find a correct generalization
candidate by the program-synthesis-based method. In contrast,
the \llm-based \proj can recognize the clamp pattern and
yield the verified generalization shown in the \cref{tab:case-study-timeout-search-space}.

\begin{table}[h]
\centering
\caption{Generalization timeout due to search-space explosion in Case Study 1.
\texttt{smin} and \texttt{smax} are the signed minimum and maximum operations.
}
\footnotesize
\begin{tabular}{l l}

\toprule

\textbf{Item} & \textbf{Optimization} \\

\midrule

\textbf{(a)} Original optimization &

$\begin{aligned}
  & \smin\left(\smax(x - 60, -155), 100\right) = x - 60\\
  \Rightarrow \; & (x + 95) \ltUnsigned 256
\end{aligned}$ \\

\midrule

\textbf{(b)} \hydra generalization & Timeout \\

\midrule

\textbf{(c)} \proj generalization &

$\begin{aligned}
& C_2 \leSigned C_3 \land (C_3 - C_2) \neq -1 \\
\vDash \; & \smin\left(\smax(x + C_1, C_2), C_3\right) = x + C_1\\
\Rightarrow \; & (x + (C_1 - C_2)) \ltUnsigned (C_3 - C_2 + 1)
\end{aligned}$ \\

\bottomrule

\end{tabular}
\label{tab:case-study-timeout-search-space}
\end{table}

\myparagraph{Case Study 2: \ungen}
\cref{tab:case-study-under-generalization} shows
a peephole optimization instance~\cite{llvm_issue_55739}
where \hydra produced an under-generalized rewrite.
In this example, \proj recovered a semantic characterization
of the bit-position test: under the \precondition that $C_1$
and $C_2$ are powers of two, the guard $(C_2 \andOp (C_1 \ll x))
\neq 0$ holds precisely when $x = \cttz(C_2) - \cttz(C_1)$,
where $\cttz$ is the count-trailing-zeros operation.
In contrast,
\hydra returned a correct but instance-tied rule that hard-codes
a specific relation between constants (e.g., $C_5 = 8 \cdot C_6$)
and retains the concrete conclusion $x = 3$, thereby substantially
narrowing the applicability of the generalization. This gap
stems from \hydra's synthesis procedure: it is greedy and
time-bounded, and it terminates as soon as it finds a
candidate that passes verification.
Moreover, \hydra compares verified candidates mainly through a
limited collection of predefined dataflow facts (\eg
KnownBits), rather than through a broader semantic notion of
generality. When these facts are insufficient to distinguish
between multiple verified candidates, it becomes difficult for
\hydra to determine which generalization has wider
applicability, and it may therefore not consistently prefer
the broader one.

\begin{table}[h]
\centering
\caption{Under-generalization in Case Study 2.
\texttt{cttz} is the count-trailing-zeros operation,
which returns the number of trailing zeros in the binary representation of its argument.
}
\footnotesize
\begin{tabular}{ll}

\toprule

\textbf{Item} & \textbf{Optimization} \\

\midrule

\textbf{(a)} Original optimization &

$\begin{aligned}
& 16 \andOp (2 \ll x) \neq 0 \Rightarrow x = 3
\end{aligned}$ \\

\midrule

\textbf{(b)} \hydra generalization &

$\begin{aligned}
& \poweroftwo(C_5) \land \poweroftwo(C_6) \land C_5 = 8 \cdot C_6 \\
\vDash \; & C_5 \andOp (C_6 \ll x) \neq 0 \Rightarrow x = 3
\end{aligned}$ \\

\midrule

\textbf{(c)} \proj generalization &

$\begin{aligned}
& \poweroftwo(C_1) \land \poweroftwo(C_2) \\
\vDash \; & (C_2 \andOp (C_1 \ll x)) \neq 0 \Rightarrow x = \cttz(C_2)-\cttz(C_1)
\end{aligned}$ \\

\bottomrule

\end{tabular}
\label{tab:case-study-under-generalization}
\end{table}

\myparagraph{Case Study 3: Generalization on Floating-Point Optimizations}
~\cref{tab:case-study-float-domain-coverage} presents a representative
floating-point clamp optimization~\cite{llvm_issue_185554}.
This case shows the third key difference: \proj handles floating-point
optimizations, while \hydra does not.
Generalizing floating-point optimizations requires reasoning
under IEEE-754 semantics~\cite{ieee7542019}, where even simple
algebraic transformations may be invalidated by NaNs, $\pm\infty$,
signed zeros, or rounding, \etc
For example, \cref{tab:case-study-float-domain-coverage}(c)
includes two seemingly ``redundant'' preconditions,
$\fcmp_\oeq(-C_1 / 2 \cdot 2, -C_1)$
and $\fcmp_\oeq(C_1\cdot 0, 0)$.
However, they are necessary under IEEE-754 semantics to ensure
no underflow to signed zero in $C_1$ and that $C_1$ is finite
(avoiding NaNs from $C_1\cdot 0$).
The complexity in floating-point operation semantics enlarges
the search space relative to integer patterns, making it more
difficult for synthesis-based approaches to
synthesize a correct generalization result within a fixed
search budget.
Moreover, \hydra builds its generalization pipeline upon the
\souper superoptimizer, which only supports integer and bitvector
operations. Extending to floating-point peephole generalization
requires substantial manual engineering, including supporting
floating-point related operators, IEEE-754 semantics,
and re-tuned search heuristics, \etc
In contrast, \proj uses \llms to generate a generalization
together with the IEEE-754 precondition required for the
optimization to be valid (e.g., finiteness and lossless scaling
constraints in ~\cref{tab:case-study-float-domain-coverage}(c)).
Therefore, \proj mitigates the combinatorial blowup that arises
from enumerating candidate precondition beyond integer-only
patterns, an inherent limitation of program-synthesis-based
approaches that can incur substantial computational overhead.

\begin{table}[h]
\centering
\caption{Floating-point optimization generalization in Case Study 3.
\texttt{fcmp\_oeq} is the ordered-equal floating-point comparison.
}
\footnotesize
\begin{tabular}{ll}

\toprule

\textbf{Item} & \textbf{Optimization} \\

\midrule

\textbf{(a)} Original optimization &

$\begin{aligned}
& \fcmp_\oeq(2 \cdot x - 1, 0) \Rightarrow x = 0.5
\end{aligned}$ \\

\midrule

\textbf{(b)} \hydra generalization & Unsupported (integer-only) \\

\midrule

\textbf{(c)} \proj generalization &

$\begin{aligned}
& \fcmp_\oeq(-C_1/2 \cdot 2, -C_1) \land \fcmp_\oeq(C_1 \cdot 0, 0) \\
\vDash \; & \fcmp_\oeq(2 \cdot x + C_1, 0) \Rightarrow x = -C_1 / 2
\end{aligned}$ \\

\bottomrule

\end{tabular}
\label{tab:case-study-float-domain-coverage}
\end{table}

\finding{\proj effectively generalizes real-world peephole
optimizations, retaining \ValPeepGenOverallCount verified
generalizations out of \ValTotalBenchmark instances across
various instruction domains. On integer-related
optimizations, it not only achieves a higher success count than
\hydra (\ValPeepGenIntCount vs.\ \ValHydraIntCount), but also
successfully generalizes \emph{every} instance solved by \hydra; over this
shared comparison set, \proj more often yields a strictly more
general rule (\ValPeepVsHydraBetterCount vs.\
\ValPeepVsHydraWorseCount).
The case studies further show that \proj can overcome
search-space explosion and under-generalization, while also
extending generalization to floating-point optimizations beyond
\hydra's scope.}

\subsection{RQ2: \rqtwo}
\label{subsec:rq2}
To assess how each generalization strategy in \proj affects the
quality of the synthesized rewrite rules, we quantified the effectiveness
of each strategy by counting the number of optimization instances in
which it contributes to a retained generalization, \ie, one that
passes the verfication described in \cref{subsec:verification}
and attains broader applicability.
This analysis considers the four stages described in
~\cref{sec:methodology}.
For finer-grained analysis, we further decomposed constraint
relaxation into three substeps: removing precondition, weakening
precondition, and removing flags or attributes.

We instrumented \proj to record a strategy application as
effective when it produces a candidate rule that is retained
by \proj.
We then aggregated these effective applications across all successful
optimization instances and reported the number of accepted applications
for each strategy stage or substep under constraint relaxation.

{
\begin{table}[H]
\footnotesize
\centering
\caption{Contribution analysis of \proj's generalization strategies.
``\#Effective/\#Success'' represents effective applications over all
successfully generalized optimizations.}
\label{tab:rq2-pass-effectiveness}
\begin{tabular}{lc}
\toprule
\textbf{Strategy} & \textbf{\#Effective/\#Success} \\
\midrule
Constant Symbolic Generalization & \ValPassConstSymEffective/\ValPeepGenOverallCount \\
Structural Generalization & \ValPassStructuralEffective/\ValPeepGenOverallCount \\
Constraint Relaxation (Remove precondition) & \ValPassRemovePCEffective/\ValPeepGenOverallCount \\
Constraint Relaxation (Weaken precondition) & \ValPassWeakenPCEffective/\ValPeepGenOverallCount \\
Constraint Relaxation (Flag \& Attribute Removal) & \ValPassFlagAttributeRemovalEffective/\ValPeepGenOverallCount \\
Bitwidth/Precision Generalization & \ValPassTypeGeneralizationEffective/\ValPeepGenOverallCount \\
\bottomrule
\end{tabular}
\end{table}
}

\subsubsection{Results}

\Cref{tab:rq2-pass-effectiveness} shows that both high-level
strategy stages and finer-grained constraint-relaxation operations
make meaningful contributions to \proj's accepted generalizations.
Constant symbolic generalization and structural generalization account
for \ValPassConstSymEffective and \ValPassStructuralEffective effective
applications, respectively, while bitwidth/precision generalization
contributes \ValPassTypeGeneralizationEffective.
Within constraint relaxation, effective applications arise from
removing precondition (\ValPassRemovePCEffective), weakening
precondition (\ValPassWeakenPCEffective), and removing flags or
attributes (\ValPassFlagAttributeRemovalEffective).
Taken together, these results indicate that \proj's effectiveness
depends on all of these strategies to enlarge the applicability
of original peephole optimizations.

{
\begin{table}[H]
\centering
\caption{Change magnitude of effective strategy applications in \proj.
``\#Affected/\#Effective'' reports the number of transformed elements
per effective application. ``\#Affected'' counts the number of
generalized constants, structurally abstracted elements, removed
precondition, weakened precondition, removed flags or attributes.}
\label{tab:rq2-pass-affected}
\footnotesize
\begin{tabular}{lc}
\hline
\textbf{Strategy} & \textbf{\#Affected/\#Effective} \\
\hline
Constant Symbolic Generalization & \ValPassConstSymAffected/\ValPassConstSymEffective \\
Structural Generalization & \ValPassStructuralAffected/\ValPassStructuralEffective \\
Constraint Relaxation (Remove precondition) & \ValPassRemovePCAffected/\ValPassRemovePCEffective \\
Constraint Relaxation (Weaken precondition) & \ValPassWeakenPCAffected/\ValPassWeakenPCEffective \\
Constraint Relaxation (Flag \& Attribute Removal) & \ValPassFlagAttributeRemovalAffected/\ValPassFlagAttributeRemovalEffective \\
\hline
\end{tabular}
\end{table}
}

Table~\ref{tab:rq2-pass-affected} shows the contribution counts
by characterizing the change magnitude of each effective strategy
application. Constant symbolic generalization generalizes
\ValPassConstSymAffected constants across
\ValPassConstSymEffective effective applications, while structural
generalization abstracts \ValPassStructuralAffected structural
elements across \ValPassStructuralEffective effective applications.
Within constraint relaxation, removing precondition, weakening
precondition, and removing flags or attributes affect
\ValPassRemovePCAffected, \ValPassWeakenPCAffected, and
\ValPassFlagAttributeRemovalAffected elements over
\ValPassRemovePCEffective, \ValPassWeakenPCEffective, and
\ValPassFlagAttributeRemovalEffective effective applications,
respectively.
For bitwidth/precision generalization, we did not count the
number of affected elements, it is applied to an optimization
as a whole rather than to individual elements.
Taken together, these results show that \proj's strategies
differ not only in how often they contribute, but also in the
granularity of the transformations they introduce.

\finding{All strategy stages contribute to \proj's final
effectiveness, indicating that high-quality generalization
depends on their complementarity rather than any single
transformation. The results further show that these strategies
operate at different dimensions, and collectively enlarge the
applicability of generalization rules}

\subsection{RQ3: \rqthree}
\label{subsec:rq3}
To answer RQ3, we instantiated \proj with four Qwen3 variants:
\QBI, \QBT,
\QST, and \QSI. These models
span two model scales (30B and 235B) and two
modes (Instruct and Thinking), where Instruct variants generate
responses by directly following the prompt, whereas Thinking
variants perform explicit reasoning before producing an answer.
Therefore, it enables a controlled
comparison of how model capacity and explicit reasoning behavior
affect \proj's ability to generalize real-world peephole
optimizations. To minimize the confounding effect of potential
data leakage, we conducted this comparison exclusively on
\projdataset. For each model, we ran the same preprocessing,
prompting, verification, and post-processing pipeline on the
same \projdataset instances, so that the only changing factor
is the underlying \llm. In particular, comparing \QBI with
\QSI isolates the effect of model scale under
the same non-reasoning interface, while comparing
\QBT with \QST provides the
corresponding scale comparison under reasoning mode. Likewise,
comparing the Thinking and Instruct variants at the same scale
allows us to assess the marginal effect of enabling explicit
reasoning. Moreover, all four Qwen3 variants share the same
knowledge cutoff date (April 2025), which helps reduce confounding
effects from unequal training horizons.

{
\begin{table}[H]
\centering
\footnotesize
\caption{Model comparison for RQ3. }
\label{tab:rq3-model-comparison}
\begin{tabular}{lcccc}
\toprule
\textbf{Model} & \textbf{Reasoning} & \textbf{\#Success}/\textbf{\#Total} \\
\midrule
\QBI & No & \ValRQThreeQwenTwoThirtyFiveInstructSuccessCount/\ValprojBenchmark \\
\QBT & Yes & \ValRQThreeQwenTwoThirtyFiveThinkingSuccessCount/\ValprojBenchmark \\
\QSI & No & \ValRQThreeQwenThirtyInstructSuccessCount/\ValprojBenchmark \\
\QST & Yes & \ValRQThreeQwenThirtyThinkingSuccessCount/\ValprojBenchmark \\
\bottomrule
\end{tabular}
\end{table}
}

\subsubsection{Results}

Table~\ref{tab:rq3-model-comparison} reports the model comparison
results on \projdataset. Among the four variants, \QBT achieves
the strongest performance, retaining
\ValRQThreeQwenTwoThirtyFiveThinkingSuccessCount out of
\ValprojBenchmark generalizations. It is followed by \QST with
\ValRQThreeQwenThirtyThinkingSuccessCount retained
generalizations, \QBI with
\ValRQThreeQwenTwoThirtyFiveInstructSuccessCount, and \QSI with
\ValRQThreeQwenThirtyInstructSuccessCount. Two main findings
emerge from these results. First, enabling explicit reasoning
consistently improves performance at both model scales. At 235B,
switching from Instruct to Thinking increases the number of
retained generalizations from
\ValRQThreeQwenTwoThirtyFiveInstructSuccessCount to
\ValRQThreeQwenTwoThirtyFiveThinkingSuccessCount; at 30B, the
same change increases the number from
\ValRQThreeQwenThirtyInstructSuccessCount to
\ValRQThreeQwenThirtyThinkingSuccessCount. This trend is
consistent with the view that the Thinking mode provides
stronger effective reasoning capability, particularly for
optimizations whose generalized rules require preconditions with
multiple conjunctive clauses. In such cases, the model appears
better able to capture the semantic structure implicit in the
original optimization and translate it into valid, more broadly
applicable rewrite rules.
Second, increasing model scale also improves performance, but
its effect depends on the reasoning mode. Under Instruct mode,
scaling from 30B to 235B yields only a modest gain, from
\ValRQThreeQwenThirtyInstructSuccessCount to
\ValRQThreeQwenTwoThirtyFiveInstructSuccessCount. In contrast,
under Thinking mode, the same scale increase leads to a much
larger improvement, from \ValRQThreeQwenThirtyThinkingSuccessCount
to \ValRQThreeQwenTwoThirtyFiveThinkingSuccessCount.

\finding{Model choice materially affects \proj's performance,
with \QBT achieving the best result
(\ValRQThreeQwenTwoThirtyFiveThinkingSuccessCount/\ValprojBenchmark).
Explicit reasoning consistently improves performance at both
scales, and the benefit of larger model scale is substantially
greater under reasoning mode.}

\subsection{Threats to Validity}
\label{subsec:threats-to-validity}
The first threat to validity is potential data leakage. Since our
task is to generalize previously reported peephole
optimizations, it is possible that the training data of the
evaluated \llms already contain information related to some
benchmark instances, which could inflate the results. To mitigate
this threat, we curated \projdataset using \llvm issues reported
after the knowledge cutoff dates of the evaluated \llms and
report its results separately from \Hydradataset. Another threat
is that the collected benchmarks may not be fully representative
of all real-world peephole optimizations. To mitigate this
threat, we evaluated \proj on two datasets covering multiple
instruction domains and retained only carefully curated cases
that passed our verification pipeline.

%% file: related-work.tex
In this section, we discuss the previous work related to \proj.

\subsection{Generalizing Peephole Optimizations}
\hydra~\cite{hydra} is the most closely related work to \proj.
It generalizes specific peephole optimizations based on program
synthesis on top of \souper~\cite{souper}. As discussed in \cref{subsec:peephole-optimization-generalizations},
\hydra remains limited by search-space
explosion and fails on complex patterns, its heuristic
and greedy synthesis strategy can yield under-generalized
result without semantic understanding, and it only supports
peephole optimizations in the integer domain.
Based on superoptimization and enumerative search, Optgen~\cite{buchwald2015optgen}
is another peephole optimization generalization work supporting
symbolic constants and preconditions.
Optgen works by enumerating all possible preconditions and
all possible constant expressions, but it does not allow users
to generalize a specific, ungeneralized optimization and it only
supports optimizations at 8 and 32 bits.
Besides, it does not support dataflow facts (\eg PowerOfTwo),
which are crucial for generalizaion.
Overall, it only supports a limited set of optimizations and
less suited to generalizing specific real-world
optimizations or more complex rewrites.
Proofgen~\cite{proofgen} follows a different route to generalizing
peephole optimizations.
It adopts a proof-centric approach that derives
generalized optimizations from explicit correctness
proofs, rather than from a generate-and-verify workflow.
This direction is powerful in principle, but it shows some limitations.
First, the generalization is strictly bound to the explicit
proof structure and different proofs or even
different linearization orders of the same proof tree can yield
entirely incomparable rules. Second, its capability is
bottlenecked by the predefined axioms, restricting the system
to concepts explicitly captured within that domain. Finally,
this proof-centric approach has not yet been shown to scale
to the complex, real-world generalizations.

In contrast, \proj takes a more flexible and scalable approach to
generalizing peephole optimizations.
\proj organizes generalization as a closed-loop pipeline
that combines \llm-guided symbolic constant generalization,
structural generalization, constraint relaxation, and
bitwidth/precision generalization with feedback from syntactic
checking, semantic verification, and profitability assessment.
This approach allows \proj to handle a much broader range of
optimization patterns, and to discover more general
optimization rules.

\subsection{Superoptimization}
Superoptimization is a more aggressive optimization technique that
searches for an equivalent but better implementation of a given program
fragment, usually with solver-backed or formally checked validation.
Massalin \etal introduced the classical
superoptimizer based on exhaustive enumeration of short instruction
sequences~\cite{massalin1987superoptimizer}. Josh \etal later proposed Denali,
formulating superoptimization as goal-directed search guided by theorem
proving~\cite{denali}, while Bansal and Aiken showed how brute-force
search can automatically build peephole
superoptimizers~\cite{peephole_superoptimizers_generation}.
Schkufza \etal proposed STOKE,
which further improved scalability by using stochastic search for loop-free
x86 code~\cite{stoke}.
Within the \llvm ecosystem, \souper synthesizes more efficient
equivalents for a restricted, dataflow-oriented subset of \llvm
\ir~\cite{souper}, and \minotaur extends this direction toward
SIMD-oriented code and broader instruction support~\cite{Minotaur}.

These systems are effective at discovering concrete rewrites, but they
generally target fixed program fragments and concrete constants. They do
not directly address our setting of starting from a concrete missed
optimization and inferring a reusable generalized rule with symbolic
constants, dataflow facts, and preconditions.

\subsection{Automatic Missed Optimization Detection}
A complementary line of work focuses on automatically exposing
optimization opportunities that production compilers miss. Barany uses
differential testing to identify missed compiler
optimizations~\cite{dt4missedopt1}. Taneja \etal use offline, SMT-based
checking to expose imprecision in static analyses that can lead to
missed optimizations~\cite{fm4missedopt1}. Theodoridis \etal propose
using dead code elimination as a lens for finding missed optimizations
in LLVM and GCC~\cite{dt4missedopt3}, and later show that providing
compilers with refined inputs can still lead to degraded optimized
output because of unexpected optimization
interactions~\cite{Theodoridis2024Refined}. DITWO extends differential
testing to WebAssembly optimizers~\cite{dt4missedopt4}, and SBC
cross-validates coordinated optimization behaviors to detect missed
opportunities across passes~\cite{dt4missedopt2}. Most recently,
\lampo combines \llms with formal verification to discover missed
peephole optimizations in LLVM~\cite{lpo}.

These techniques are complementary to \proj. They help surface missed
opportunities or the analysis imprecision that causes them, but they do
not focus on synthesizing a generalized optimization rule from a
concrete optimization instance.

\subsection{Finding Weakest Preconditions}
Another related line of work focuses on inferring
the weakest or sufficiently weak preconditions for an
already known transformation, rather than synthesizing the
transformation itself.
\psyco~\cite{psyco} is an early compiler-specific system in
this space and synthesizes weakest preconditions for compiler
optimizations within a predefined precondition language.
\aliveinfer~\cite{alive-infer} later adopts a data-driven
strategy for \llvm peephole optimizations expressed in Alive:
it learns predicates that separate positive and negative
examples, and reports both weakest and succinct partial
preconditions.
Outside compiler optimization, \pie~\cite{pie} and
\pgen~\cite{pgen} study the broader problem of program
precondition inference.
\pie learns features on demand from examples to infer richer
preconditions, while \pgen uses counterexample-guided
refinement to derive necessary and sufficient preconditions.

These techniques are complementary to \proj.
They all assume that the source and target programs are
already fixed and aim to infer only the guard under which
the optimization or program is valid, whereas \proj targets
a harder setting: starting from concrete
optimizations, it infers the generalized optimization rule
itself together with its applicable conditions.

%% file: conclusion.tex
This paper presents \proj, a novel framework for generalizing
peephole optimizations with large language models.
Our central insight is that \llms are effective at semantic
abstraction and creative reasoning during generalization,
while rigorous verification is necessary to ensure the
correctness of the generalization.
Accordingly, \proj organizes peephole generalization as a closed-loop
pipeline that combines symbolic constant generalization, structural
generalization, constraint relaxation, and bitwidth/precision
generalization with syntactic checking, formal semantic verification,
and profitability assessment.
Through a comprehensive evaluation on real-world
peephole optimizations from the \llvm ecosystem, we show that
\proj successfully generalizes \ValPeepGenOverallCount out of
\ValTotalBenchmark optimization instances across integer,
floating-point, and memory domains.
On integer-related optimizations, \proj also substantially
outperforms \hydra in both success rate and generality, demonstrating
that \llm-guided generalization can overcome the search-space
explosion, \ungen, and domain restrictions that limit prior
program-synthesis-based approaches.

Our results suggest that \llms can serve as a practical foundation
for compiler optimization when they are coupled with strong verification
rather than used as standalone generators.
The success of \proj indicates that similar methodologies
could be applied to other compiler tasks that require semantic
abstraction over concrete program transformations.
More broadly, as reasoning-capable \llms continue to improve, we
expect frameworks like \proj to become an increasingly useful tool
for helping compiler developers discover broader, more reusable, and
more reliable optimization rules for modern compilation pipelines.